\documentclass[namedreferences]{solarphysics}
\usepackage[hyperref,optionalrh,solaromanenum]{spr-sola-addons} 
\usepackage{graphicx}                                           
\usepackage{color,xcolor, colortbl}                             
\usepackage{breakurl}                                           
\usepackage{gensymb,longtable,ulem}
\hypersetup{breaklinks,colorlinks,citecolor=blue,linkcolor=blue,urlcolor=blue,pdfborder={0 0 0 }}
\usepackage{lscape}

\usepackage{booktabs}
\usepackage{pdflscape}
\usepackage{lipsum}
\usepackage{caption}
\usepackage{siunitx}

\chardef\us=`\_

\begin{document}

\begin{article}
\begin{opening}

\title{Solar observations with single-dish INAF radio telescopes: continuum imaging in the 18 -- 26~GHz range}

%
\author[addressref={oac},corref,email={alberto.pellizzoni@inaf.it}]{\inits{A.}~\fnm{A.}~\lnm{Pellizzoni}~\orcid{0000-0002-4590-0040}}
\author[addressref=ira]{\inits{S.}~\fnm{S.}~\lnm{Righini}~\orcid{0000-0001-7332-5138}}
\author[addressref=asi]{\inits{M.~N.}~\fnm{M.~N.}~\lnm{Iacolina}~\orcid{0000-0003-4564-3416}}
\author[addressref=oac]{\inits{M.}~\fnm{M.}~\lnm{Marongiu}~\orcid{0000-0002-5817-4009}}
\author[addressref={unica,asi}]{\inits{S.}~\fnm{S.}~\lnm{Mulas}~\orcid{0000-0002-5455-1233}}
\author[addressref=ext]{\inits{G.}~\fnm{G.}~\lnm{Murtas}~\orcid{0000-0002-7836-7078}}
\author[addressref=asi]{\inits{G.}~\fnm{G.}~\lnm{Valente}~\orcid{0000-0002-6503-2498}}
\author[addressref=oac]{\inits{E.}~\fnm{E.}~\lnm{Egron}~\orcid{0000-0002-1532-4142}}
\author[addressref=oac]{\inits{M.}~\fnm{M.}~\lnm{Bachetti}~\orcid{0000-0002-4576-9337}}
\author[addressref=oac]{\inits{F.}~\fnm{F.}~\lnm{Buffa}~\orcid{0000-0001-9256-4476}}
\author[addressref=oac]{\inits{R.}~\fnm{R.}~\lnm{Concu}~\orcid{0000-0003-3621-349X}}
\author[addressref=oac]{\inits{G.~L.}~\fnm{G.~L.}~\lnm{Deiana}~\orcid{0000-0002-5404-5162}}
\author[addressref=oact]{\inits{S.~L.}~\fnm{S.~L.}~\lnm{Guglielmino}~\orcid{0000-0002-1837-2262}}
\author[addressref=oac]{\inits{A.}~\fnm{A.}~\lnm{Ladu}~\orcid{0000-0003-1920-9560}}
\author[addressref=oact]{\inits{S.}~\fnm{S.}~\lnm{Loru}~\orcid{0000-0001-5126-1719}}
\author[addressref=ira]{\inits{A.}~\fnm{A.}~\lnm{Maccaferri}~\orcid{0000-0001-7231-4007}}
\author[addressref=oac]{\inits{P.}~\fnm{P.}~\lnm{Marongiu}~\orcid{0000-0003-0314-7801}}
\author[addressref=oac]{\inits{A.}~\fnm{A.}~\lnm{Melis}~\orcid{0000-0002-6558-1315}}
\author[addressref=oac]{\inits{A.}~\fnm{A.}~\lnm{Navarrini}~\orcid{0000-0002-6191-6958}}
\author[addressref=ira]{\inits{A.}~\fnm{A.}~\lnm{Orfei}~\orcid{0000-0002-8723-5093}}
\author[addressref=oac]{\inits{P.}~\fnm{P.}~\lnm{Ortu}~\orcid{0000-0002-2644-2988}}
\author[addressref=oac]{\inits{M.}~\fnm{M.}~\lnm{Pili}~\orcid{0000-0003-3715-1091}}
\author[addressref=oac]{\inits{T.}~\fnm{T.}~\lnm{Pisanu}~\orcid{0000-0003-2510-7501}}
\author[addressref=ira]{\inits{G.}~\fnm{G.}~\lnm{Pupillo}~\orcid{0000-0003-2172-1336}}
\author[addressref=asi]{\inits{A.}~\fnm{A.}~\lnm{Saba}~\orcid{0000-0002-1607-5010}}
\author[addressref=oac]{\inits{L.}~\fnm{L.}~\lnm{Schirru}~\orcid{0000-0002-8199-6510}}
\author[addressref=asi]{\inits{G.}~\fnm{G.}~\lnm{Serra}~\orcid{0000-0003-0720-042X}}
\author[addressref={oac,astron}]{\inits{C.}~\fnm{C.}~\lnm{Tiburzi}~\orcid{0000-0001-6651-4811}}
\author[addressref=ira]{\inits{A.}~\fnm{A.}~\lnm{Zanichelli}~\orcid{0000-0002-2893-023X}}
\author[addressref=astron]{\inits{P.}~\fnm{P.}~\lnm{Zucca}~\orcid{0000-0002-6760-797X}}
\author[addressref={oats,units}]{\inits{M.}~\fnm{M.}~\lnm{Messerotti}~\orcid{0000-0002-5422-1963}}

%
\address[id=oac]{INAF - Cagliari Astronomical Observatory, Via della Scienza 5, I--09047 Selargius (CA), Italy}
\address[id=ira]{INAF - Institute of Radio Astronomy, Via Gobetti 101, I--40129 Bologna, Italy}
\address[id=asi]{ASI - c/o Cagliari Astronomical Observatory, Via della Scienza 5, I--09047 Selargius (CA), Italy}
\address[id=unica]{Department of Physics, University of Cagliari, SP Monserrato-Sestu, KM 0.7, I--09042 Monserrato (CA), Italy}
\address[id=ext]{College of Engineering, Mathematics and Physical Sciences,Harrison Building, Streatham Campus, University of Exeter, North Park Road, Exeter, EX4 4QF, UK}
\address[id=oact]{INAF - Catania Astrophysical Observatory, Via Santa Sofia 78, I--95123 Catania, Italy}
\address[id=astron]{ASTRON – The Netherlands Institute for Radio Astronomy, Oude Hoogeveensedijk 4, 7991 PD Dwingeloo, The Netherlands}
\address[id=oats]{INAF – Trieste Astronomical Observatory, Via Giambattista Tiepolo 11, I--34131 Trieste, Italy}
\address[id=units]{Department of Physics, University of Trieste, Via Alfonso Valerio 2, I--34127 Trieste, Italy}

%
\runningauthor{Pellizzoni et al.}
\runningtitle{Single-dish solar observations with INAF radio telescopes}

\begin{abstract}

We present a new solar radio imaging system implemented through the upgrade of the large single-dish telescopes of the Italian National Institute for Astrophysics (INAF), not originally conceived for solar observations.

During the development and early science phase of the project (2018--2020), we obtained about 170 maps of the entire solar disk in the 18--26~GHz band, filling the observational gap in the field of solar imaging at these frequencies.
These solar images have typical resolutions in the 0.7--2~arcmin range and a brightness temperature sensitivity $<$10~K.
Accurate calibration adopting the Supernova Remnant Cas~A as a flux reference, provided typical errors $<$3\% for the estimation of the quiet-Sun level components and for active regions flux measurements.  

As a first early science result of the project, we present a catalog of radio continuum solar imaging observations with Medicina 32-m and SRT 64-m radio telescopes including the multi-wavelength identification of active regions, their brightness and spectral characterization.
The interpretation of the observed emission as thermal bremsstrahlung components combined with gyro-magnetic variable emission pave the way to the use of our system for long-term monitoring of the Sun.
We also discuss useful outcomes both for solar physics (e.g. study of the chromospheric network dynamics) and space weather applications (e.g. flare precursors studies).

\end{abstract}

%
\keywords{Instrumentation and data management · Chromosphere · Sun, radio emission · Sun, corona }

\end{opening}

%

\newpage

\section{Introduction}
\label{par:intro}

The study of the Sun atmosphere through ground-based and space monitoring programs at different wavelengths represents a major topic of present astrophysics.
However, some radio frequency bands surprisingly lack solar observations, despite being relevant for the investigation of many astronomical and astrophysical phenomena at large.

Plasma processes -- such as magnetic reconnection, shocks and particle acceleration (e.g., \citealp{Alissandrakis94, Gary96, Alissandrakis97, Shibasaki98, Shibasaki11, Messerotti16, Nindos20,Alissandrakis20}) -- contribute to a complex and not fully understood radio emission picture. In this context, spatially-resolved and time-resolved radio data are crucial to complement the wealth of existing information in the UV, Optical, IR and X-ray domains.
Compared to the emission in other spectral ranges (e.g. the EUV), the radio Sun has the advantage of being understood as originating mostly from thermal bremsstrahlung in local thermodynamic equilibrium, with the addition of sporadic and variable gyro-magnetic emission in active regions (ARs; \citealp{Dulk85}).
The radio Sun can thus be used as a powerful workbench to shed light on the problem of the heating of the chromosphere and the corona (see e.g., \citealp{Shibasaki11,Alissandrakis20}).

A growing number of existing facilities\footnote{For a list see the Community of European Solar Radio Astronomers (CESRA): \url{http://www.astro.gla.ac.uk/users/eduard/cesra/?page_id=187}} -- that regularly provide solar imaging radio data -- is working on the puzzling disentanglement of the large variety of phenomena occurring in the solar atmosphere on a wide frequency range, together with non-dedicated facilities offering more sporadic solar observations (\citealp{Carley20}).
As the opacity increases with the observing wavelength, the effective height of formation moves from the temperature minimum region to the low corona. The former is visible at sub-millimeter waves and approached by several instruments, e.g. ALMA observations (\citealp{Loukitcheva19}), while the latter is mostly visible at meter waves, for example with LOFAR \citep{Zucca18,Zhang20}, the Nan{\c{c}}ay Radioheliograph (NRH, \citealp{Kerdraon97}) and GRAPH (\citealp{Ramesh98}).
In between these extreme wavelengths, the chromosphere is visible at moderately high radio-frequencies (10--50~GHz) and it hosts most of the active solar features linking the photosphere to the corona, with a temperature rise of still unclear origin, being explored by a number of dedicated facilities (see e.g. Nobeyama Radioheliograph, NoRH, \citealp{Shibasaki98}; Mets{\"{a}}hovi Radio Observatory, MRO, \citealp{Kallunki17}; the Expanded Owens Valley Solar Array, EOVSA\footnote{\url{http://www.ovsa.njit.edu/}}, \citealp{Gary18}; RATAN-600, \citealp{Schwartz78}; SSRT, \citealp{Smolkov86,Grechnev03}; MUSER, \citealp{Cheng19}).
In particular, solar atmospheric models could be critically constrained by observations in K-band (18--27~GHz) and above (see e.g. \citealp{Selhorst05,Gopalswamy16}). This portion of the spectrum is suitable for detailed measurements of the chromospheric brightness temperature of the quiet-Sun (QS) and to assess the rich thermal vs. non-thermal content of AR emission.
At these frequencies both the QS network and gyro-magnetic variable components can be simultaneously mapped due to the relatively low instrumental dynamic range required for the observations.

For such scientific applications, single-dish radio mapping of the solar disk is as well suitable as interferometric observations in K-band (see e.g. \citealp{White17,Pellizzoni19}).
Synthesis images of the full solar disk cannot be easily obtained in the frequency range 10--30~GHz through interferometric networks aimed at simultaneously resolving both large and arcsec-level solar features (see e.g. \citealp{Wilson13}), with the exception of dedicated short-baseline facilities, as the NoRH \citep{Shibasaki98} that ceased operations in 2020 \citep{masuda19}.
On the other hand, at the expenses of a coarser spatial resolution, single-dish radio mapping on relatively large and bright sources (as for example Supernova Remnants and the solar disk) offers accurate calibrated images independently from the target size, without synthesis imaging artifacts (see \citealp{Loru19,Pellizzoni19,Marongiu20,Loru21}).

Accurate measurements of the brightness temperature of the QS component are lacking in the poorly known 20--26~GHz range.
Present information suggests a possible spectral change (flattening?) in the quiet solar chromosphere at these frequencies \citep{Landi08}, but the available measurements are scattered due to the difficulty to separate the pure QS component from the contribution by the ARs.

Disentanglement and quantification of the thermal (free-free and gyro-resonance) and non-thermal emission components is an open issue for radio ARs (see e.g. \citealp{Lee07,White04}), as well as for peculiar large-scale structures such as coronal holes, loop systems, filaments, streamers and the coronal plateau, that would benefit from systematic long-term observations.
Multi-wavelength observations could provide insights on the formation process of the ARs hosting major flares and determining the slowly varying component of the Sun over the 11-year cycle.
In particular, observations of strong gyro-magnetic emission over sunspots groups can be used to test in detail the magnetic field model extrapolated from measurements at the photosphere (see \citealp{Lee98}).
In this frame, it is important to enlarge the sample of measurements of intensity, spectra and size of the bright features in coincidence with the chromospheric network as a function of the frequency.

In addition, significant spectral variations and/or the formation of a local maximum in the microwave spectrum of solar ARs appear to be an important factor in predicting powerful flares (within 24--48~hours from the event), as suggested by RATAN-600 observations (see e.g. \citealp{Borovik12}).
Radio monitoring campaigns, focusing on at least two frequencies at the low/high edge of the K-band, might provide spectral index estimates for the ARs, and detect anomalous spectral variations (e.g. sudden flattening) as a signature anticipating the flaring process.

On longer time scales, the total K-band radio flux integrated over the whole disk is known to be an excellent index of solar activity.
As suggested by \cite{Shibasaki98} radio butterfly diagrams show very well not only the migration of ARs toward the equator, but also the long-term behavior of polar brightening.
\cite{Selhorst14} demonstrated that the statistics of the number of active regions (NAR) observed at 17~GHz with NoRH \citep{Nakajima94}, and used as solar activity index, is sensitive to magnetic fields weaker than those necessary to form sunspots.
NAR minima are shorter than those of the sunspot number (SSN) and other activity index.
This could reflect the presence of ARs generated by faint magnetic fields or spotless regions, which are a considerable fraction of the counted ARs.
It is thus important that such observations continue for the decades to come (see e.g. \citealp{Zucca17,Matamoros17}).

The INAF single-dish radio telescopes network\footnote{\url{https://www.radiotelescopes.inaf.it}} includes the 64-m Sardinia Radio Telescope (SRT) and two 32-m antennas at Medicina and Noto.
They are open to the scientific community for observing time applications (both single-dish and interferometric modes) granted through call for proposals.
SRT is also operated by the Italian Space Agency (ASI) for spacecraft tracking and space science (Sardinia Deep Space Antenna, SDSA; \citealp{Parca17}).
These facilities were not originally intended and designed for solar observations.
Starting from early 2018 our group developed imaging configurations for solar observations with the 32-m Medicina and the 64-m SRT dishes in the 18--26~GHz frequency range (SunDish project, in collaboration with INAF and ASI; \citealp{Pellizzoni19,Plainaki20}).
Two years (2018--2020) of experimental solar observations on about a weekly basis, helped to establish the Italian radio telescope network as a non-dedicated solar imaging facility, 
filling a frequency gap that presently exist in the worldwide solar monitoring scenario.

In this paper we describe the implementation of the instrumental configurations for radio-continuum solar imaging and observing techniques adopted for the INAF radio-telescopes (Sect.~\ref{sec:obs_setup}); the peculiar data processing system developed \textit{ad hoc} for the production of our single-dish solar maps (Sect.~\ref{sec:data_an}); the data analysis of the first release of our solar catalog at 18--26~GHz (Sect.~\ref{par:obs_summ}) including an estimation of the QS flux levels (Sect.~\ref{subsec:qs_fluxes}) and the AR identification and their spectra in this frequency range (Sect.~\ref{subsec:ar_fluxes}).
A discussion of our early scientific results is described in Sect.~\ref{par:disc}, and finally in Sect.~\ref{sect:prosp} we give our conclusions, including future prospects towards a full implementation of solar observing modes with INAF radio telescopes.

\section{Solar instrumentation and setup}
\label{sec:obs_setup}

Our project is based on the exploitation of INAF radio telescopes for solar observations through single-dish mode operations.
These instruments were not originally designed for solar observations; even pointing them close to the Sun position was considered a potentially hazardous action for opto-mechanics and front-end thermal/electromagnetic safety.
They required additional setups and accurate electromagnetic and thermal assessments in order to guarantee instrumentation safety and to ensure a linear response to the strong solar input signal impacting on the systems amplification chains.
In K-band the solar brightness ($\sim$ 5,000~Jy/arcmin$^2$) is over three orders of magnitude higher than typical radio-astronomical calibration sources.
The implementation of variable filters for additional signal attenuation, included in the receivers amplification chain, was crucial to avoid electronic saturation in signal response and possible instrumental damage.
After equipping the telescopes with a suitable signal attenuation setup, the typical variability of solar phenomenology at these frequencies is compatible with the available instrumental dynamic range \citep{Pellizzoni19,Iacolina19}.
In fact, apart from strong flare episodes, most of the enhanced emission associated to ARs in the 18--26~GHz range does not typically exceed the QS brightness level by orders of magnitudes, as it happens instead at lower radio frequencies in the MHz-GHz domain.
In the time frame 2018--2020, most of our solar mapping in K-band were performed by the Medicina radio telescope, with a $\sim$ 10\% of the total number of sessions provided by the larger SRT.

\subsection{Medicina Radio Telescope}
\label{subsec:med}

The 32-m Medicina radio telescope operates with multiple receivers covering the range 1.3--26.5~GHz. These receivers can observe the targets one at a time, with a quick frequency switch, when necessary.
In recent years a new control system (DISCOS\footnote{\url{https://discos.readthedocs.io/en/latest/user/index.html}}) was designed and produced by INAF in order to enhance the performance of this telescope -- originally conceived for VLBI observations -- as a single-dish instrument.
Fast On-The-Fly (OTF) mapping is now one of the built-in observing modes; its sky coverage efficiency is doubled in the case of the K-band receiver, as it is a dual-feed device.
The overall K-band observable portion is 18--26.5~GHz, with an instantaneous configurable bandwidth of 2~GHz maximum.
The output of each feed consists in two separate lines: LCP (Left Circular Polarization) and RCP (Right Circular Polarization) that reach an analog total-power back-end.
The frequency and bandwidth actually employed may slightly vary with time, according to the incidence of RFI (Radio Frequency Interference).

Our project exploits the Equatorial (RA-Dec) OTF mapping technique \citep{Prandoni17}.
Observing schedules are produced using a custom generator in turn relying on solar ephemeris computed with the NASA JPL Horizons web tool\footnote{\url{ https://ssd.jpl.nasa.gov/horizons.cgi}}, as the observation of Solar system targets is currently not offered among the built-in options of the Medicina control system.
Most of the maps are performed by setting the two feeds in different dynamic ranges, so as to coevally acquire data on the bright solar disk and on the much fainter emission near the limb.
This is accomplished via a double layer of variable signal attenuation devices. 
In order to acquire spectral information, observations are carried out at the boundaries of the available radio frequency band: maps are acquired first at 18~GHz, then at 26~GHz.
The two maps are obtained by scanning the desired area in Right Ascension (RA) and Declination (Dec), respectively.
This choice reflects the need to reduce the total observing time, while permitting to investigate weak disk features that, in a single map, might be altered -- or even created -- by striping effects, usually due to fast-variable weather conditions differently affecting the many sub-scans composing the map (RFI might also contribute to this issue).
Table~\ref{tab:info_med} describes the mapping parameters and setup configurations used for the receiver and back-end in the vast majority of the observing sessions (except for periodic maintenance operations in order -e.g.- to  check for frequency bands relatively free from strong RFI).

Flux density calibration is achieved through the observation of calibration sources, either in cross-scan or mapping mode.
Mostly, the Supernova Remnant Cas~A \citep{Vinyaikin14} is observed producing 40 $\times$ 40~arcmin maps, with 2.0~scan/beamsize, in order to obtain the conversion factor from arbitrary counts to flux density units, expressed in units of Jy (see Sect.~\ref{subsec:im_calib} for calibration procedure details).
The estimate of the atmospheric opacity -- one for each frequency -- is possible thanks to the Skydip observing mode, that measures the sky brightness at a range of elevations. Skydip acquisitions require additional 12~minutes and complete the data set.

\begin{table}
\caption{OTF mapping parameters and receiver/back-end configurations for observations with the Medicina radio telescope.}
\label{tab:info_med}
\begin{tabular}{l|cc}       
\hline                      
OTF parameter               & Value                  &                     \\
\hline
Map dimensions              & $80 \times 80$~arcmin  &                     \\
Scanning speed              & $6$~arcmin/s           &                     \\
Scan interleave             & $4$~scan/beamsize      &                     \\
Scan direction              & RA for 18~GHz map      &                     \\
                            & Dec for 26~GHz map     &                     \\ 
Overall duration            & 2h 30m$^a$             &                     \\
\hline
Receiver/Back-end parameter & 18-GHz map             & 26-GHz map          \\
\hline
Frequency range (GHz)       & 18.20 -- 18.45$^b$         & 25.70 -- 25.95$^d$  \\
                            &                  18.00 -- 18.25$^c$       & 26.00 -- 26.25$^e$  \\
                            &                        & (23.50 -- 23.75)$^f$ \\
Beamsize (arcmin)           & 2.1                    & 1.5                 \\
Sampling interval (ms)      & 40                     & 40                  \\
Solar disk map on           & Feed~1                 & Feed~1              \\
Coronal map on              & Feed~0                 & Feed~0              \\
\hline
\multicolumn{3}{l}{\footnotesize\itshape $^a$ To acquire both maps} \\
\multicolumn{3}{l}{\footnotesize\itshape $^b$ 
Since 17-Jun-2018
} \\
\multicolumn{3}{l}{\footnotesize\itshape $^c$ 
Until 10-Jun-2018
} \\
\multicolumn{3}{l}{\footnotesize\itshape $^d$ 
Since 12-Jun-2019
} \\
\multicolumn{3}{l}{\footnotesize\itshape $^e$ 
Until 6-Dec-2018
} \\
\multicolumn{3}{l}{\footnotesize\itshape $^f$ Frequency range used in early observations} \\
\end{tabular}
\end{table}

\subsection{Sardinia Radio Telescope}
\label{subsec:srt}

The 64-m Sardinia Radio Telescope (SRT) presently provides solar imaging in the 18--26~GHz range by the 7~feeds dual polarization K-band receiver \citep{Bolli15,Prandoni17}, and up to 100~GHz in perspective with the planned system upgrade\footnote{\url{https://sites.google.com/a/inaf.it/pon-srt/home}} for the whole INAF network.
The radio signal is processed through full-stokes spectral-polarimetric ROACH2-based back-end (SARDARA system, 1.5~GHz bandwidth, \citealp{Melis18}).

Solar imaging observations at SRT require special hardware configuration in order to attenuate the strong solar signal in the amplification chain.
At present, the need for a manual setup of additional hardware (10~dB attenuation for the 14~multi-feed chains, 7 for each polarization section) at each solar observing session restricts the use of SRT to a few solar sessions/year.
A remote-controlled attenuation level setup is under development and it will permit a quick and smart switching from standard operations to solar observing mode ("solar agility") allowing more frequent observations.
The high dynamic range of the {\sc SARDARA} spectro-polarimeter \citep{Melis18} allows us to detect both the bright solar disk (chromospheric emission) and the weaker 
emission near the limb, in the same image.
The adopted observing technique is similar to that in use with Medicina: OTF scans providing full solar mapping (both radio telescopes use the {\sc DISCOS} antenna control system); with OTF scans we spanned the region on and around the source, covered by all the 7 feeds of the receiver, along the RA direction.
Maps with SRT were performed at 18.8 and 24.7~GHz\footnote{After a specific test phase, we decreased the frequency at 24.7~GHz (from the initial 25.5~GHz) to match better receiver performances and avoid RFI.}, near the edge of the K-band in order to minimize errors in spectral index measurements.
OTF parameters and configuration parameters for front-end and back-end are listed in Table~\ref{tab:info_SRT}.
\begin{table}
\caption{OTF mapping parameters and receiver/back-end configurations for SRT observations. All the map scans were performed along the Right Ascension direction.}
\label{tab:info_SRT}
\begin{tabular}{lccc}        
\hline                       
OTF parameter                & Value                  &              &              \\
\hline
Map dimensions               & $90 \times 90$~arcmin  &              &              \\
Scanning speed               & $6$~arcmin/s           &              &              \\
Scan interleave              & $2$~scan/beamsize$^a$  &              &              \\
Single Map duration          & 1h 45m                 &              &              \\
\hline                       
Receiver/Back-end parameter  & 18-GHz map 	          & 24-GHz map   & 25-GHz map   \\
\hline
Frequency range (GHz)        & 18.1 -- 19.5           & 24.0 -- 25.4 & 25.1 -- 26.5$^b$ \\
Beamsize (arcmin)            & 1.02     	          & 0.78  	     & 0.75         \\
Sampling interval (ms)       & 20                     & 20           & 20           \\
Number of frequency channels & 1024                   & 1024         & 1024         \\
\hline
\multicolumn{4}{l}{\footnotesize\itshape $^a$ $\times$ 7 feeds} \\
\multicolumn{4}{l}{\footnotesize\itshape $^b$ in a few sessions we adopted a reduced bandwith of 0.9 GHz} \\
\end{tabular}
\end{table}

\subsection{Observing plan and strategy}
\label{subsec:obs_plan}

The Medicina 32~-m and SRT 64~-m radio telescopes are facilities open to the scientific community.
Early radio observations related to the SunDish project relied on observing time granted through INAF Guest Observer programs, Discretionary Director Time (DDT) and instrument development and testing activities (depending on the specific task/operation).

Features on the solar disk can evolve on very short timescales, thus a daily or even a more frequent monitoring is the optimal and typical choice for dedicated radio solar facilities, as for example NoRH \citep{Shibasaki13} and MRO \citep{Kallunki20}.
SRT and Medicina perform an almost-weekly monitoring campaign.
Accounting for the bright radio ARs, which typically persist from a few days to 10--15~days, and the radio telescopes schedule constraints due to high-priority projects (such as VLBI sessions), the weekly monitoring is a good compromise for typical solar sessions with non-dedicated solar instruments in this early science phase of the project (2018--2020).
A summary of the solar observations acquired with the INAF radio telescopes during the development and early science phase of the project is given in the Appendix (Table~\ref{T:obs_summ}).
In this first paper we focus on total intensity radio continuum data.
The exploitation of full Stokes radio polarimetric data will be addressed in a forthcoming dedicated work.

\section{Data Processing}
\label{sec:data_an} 

Our scientific data products rely on calibrated brightness images of the Sun atmosphere in K-band,
centered at the solar centroid ephemeris.
Image production and calibration are performed using the SRT Single-Dish Imager (SDI), which is an IDL (Interactive Data Language\footnote{\url{https://www.l3harrisgeospatial.com/Software-Technology/IDL}}) tool designed to perform continuum and spectro-polarimetric imaging, optimized for OTF scan mapping, and suitable for most receivers/back-ends available for INAF radio telescopes (see details
and applications in \citealp{Egron17,Loru19,Pellizzoni19,Marongiu20,Loru21,Marongiu22}).
SDI generates output FITS images suited to further analysis by standard astronomy tools.
The core of our procedure is to fully exploit the availability of a significant number of measurements per beam (higher than Nyquist requirement) that allows us to optimize the spatial resolution with pixel size typically chosen to be about 1/4 of the Half Power Beam Width (HPBW).
This oversampling permits to have a straightforward evaluation of statistical errors (through standard deviation of the measurements in each pixel), efficient RFI outliers removal and accurate background baseline subtraction.

A new Python package version designed for the quicklook, imaging and analysis of single-dish radio data with SRT, called SRT Single Dish Tools (SDT), is also publicly available for data processing\footnote{\url{https://srt-single-dish-tools.readthedocs.io/en/latest/}}.
We used it as a cross-check for data analysis.

The major steps/procedures involved in our data processing pipeline are described in the following sections, while the tools used for scientific data analysis are reported in Sect.~\ref{par:obs_summ}; for further information about our solar pipeline, see \citet{Marongiu22}.

\subsection{Radio Frequency Interference rejection}
\label{subsec:rfi_rejection}

A spectral RFI 'flagging', based on automated search for outliers in each scan-sample spectrum, is available for SRT when observing with the spectro-polarimetric back-end.
This has been proved to be an effective method for RFI removal (see e.g. \citealp{Loru21}).
However, such automated search is not available when observing with Total Power/Intensity back-ends, such as that in use at Medicina.
Therefore, for most of our observations, we provide an alternative automated ‘spatial RFI flagging’ procedure that consists in splitting the map into sub-regions, which correspond to adjacent solid angles in the sky.
These areas have to be inferior to the beam size (typically 1/4--1/5 of HPBW) in order to avoid discarding actual fluctuations from the source, but large enough so that they include a significant (typically $>$10) number of measurements.
The ‘outlying’ samples presenting a count level above a standard deviation-based threshold (typically $5\sigma$ level above average) are then flagged as RFI.
An additional manual/interactive RFI flagging process is then applied to remove residual artifacts.
Through the above procedures, we were able to remove most of the RFI features in the images (e.g. stripes due to commercial ground-based or satellite radio links) affecting a growing fraction of our data.

\subsection{Baseline background subtraction}
\label{subsec:baseline}

Automated baseline subtraction of radio background is performed scan by scan by SDI, using different methods adapted to each specific imaging target (see \citealp{Egron17,Loru19,Loru21}).
Since weak solar emission is still detected over 2~deg from the Sun centroid (well outside our typical mapping sizes), we could in principle model the outer brightness profiles through polynomial functions assuming the asymptotic value of such a fit as the background baseline.
In fact, even a naive linear approximation of the baseline trivially taken by connecting the minimum values at the beginning and at the end of the scan, yield negligible discrepancies ($<$0.1\% errors) in the image sensitivity, with respect to the above more complicated procedure.
We adopted this simpler yet robust method in our data processing pipeline 
for the analysis of the bright solar disk emission.

\subsection{Image production}
\label{subsec:im_production}

The control system in use at the INAF radio telescopes provides celestial coordinates for each measurement performed during the OTF scans. 
These scan measurements are then binned through an ARC-tangent projection using pixel sizes of about 1/4 of the HPBW, which corresponds to the effective resolution of the images.
Bright and adjacent point-like image features (e.g. having $> 0.1$~Jy flux density and a beam-size separation) associated with a Gaussian Point Spread Function (i.e. the antenna beam shape) are not distinguishable when the image pixel size is equal to the HPBW, while these are resolved when adopting a pixel size equal to the effective resolution (about 1/4 $\sim$ HPBW).
This arises from Gaussian beam oversampling in our mapping procedures.

\begin{figure}
\centering
{\includegraphics[width=120mm]{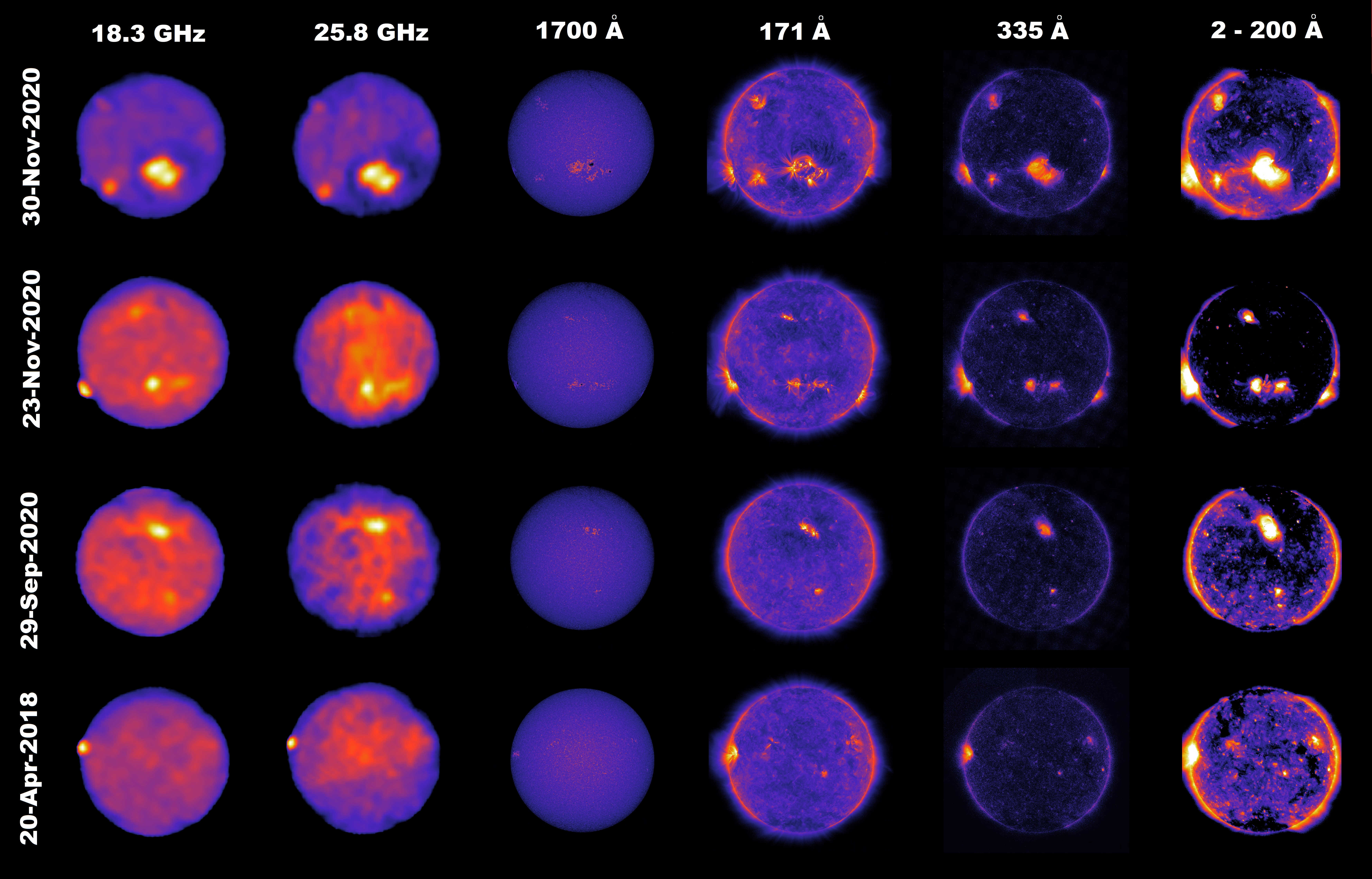}} \quad
{\includegraphics[width=120mm]{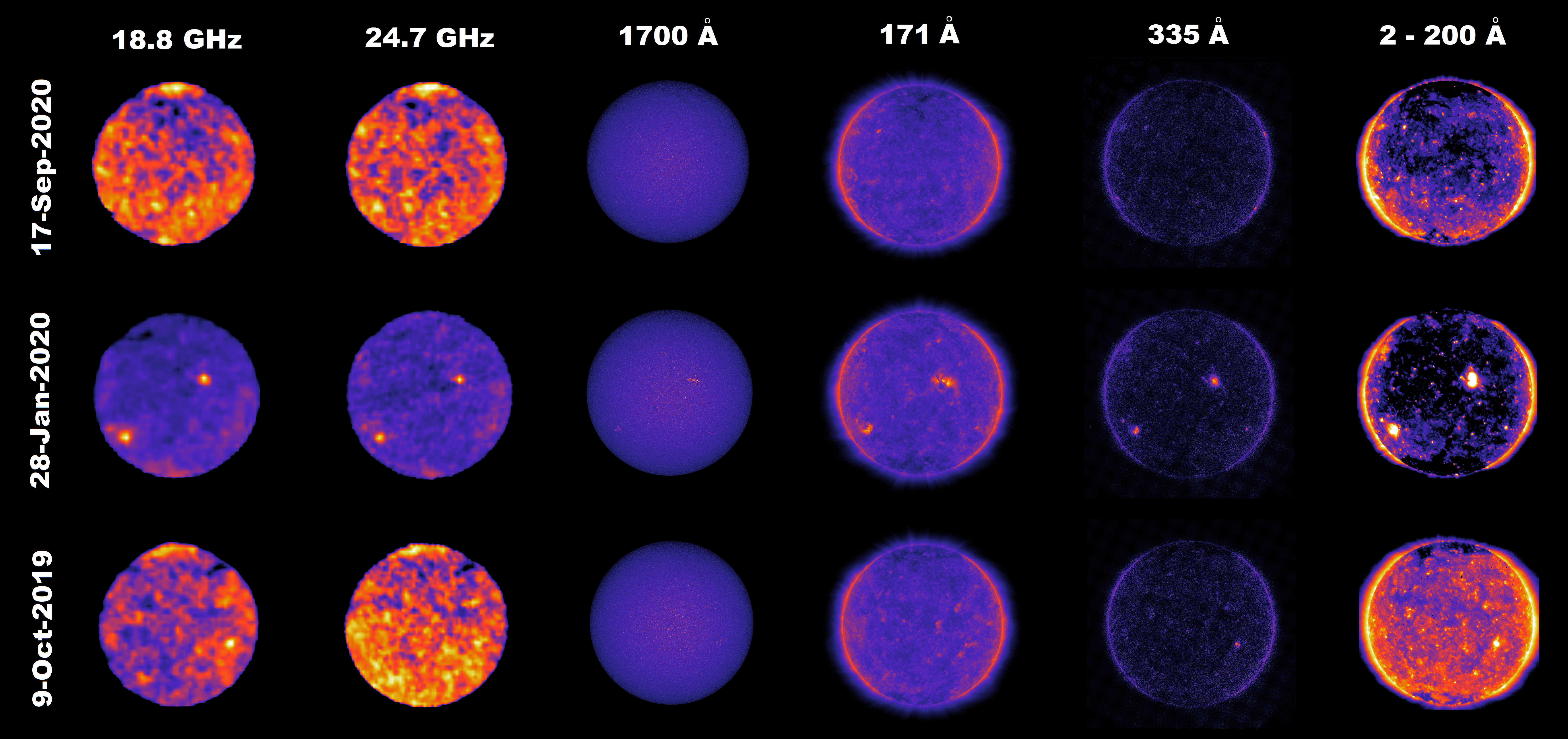}} \\
\caption{Examples of solar disk maps collected at different frequencies with Medicina (top) and SRT (bottom), in comparison with UV/EUV (SDO/AIA, \citealp{Lemen12}) and X-ray (Hinode/XRT, \citealp{Kosugi07,Golub07}) images (Credits: \href{https://sdo.gsfc.nasa.gov/}{NASA}/\href{https://www.isas.jaxa.jp/home/solar/}{JAXA}).
ARs and disk structures are clearly detected in the radio images allowing multi-wavelength spectral analysis.
}
\label{fig:overview_images}
\end{figure}

Since the apparent proper motion of the Sun in celestial coordinates is about 2.5'/hour, a blurring effect comparable to the beam size is affecting the raw images for typical mapping time of about 1.5~hours.
In order to obtain corrected astrometric images, we subtracted the actual coordinates of the Sun centroid from the celestial coordinates of each OTF sample.
We adopted the NASA/JPL Sun ephemeris\footnote{\url{https://ssd.jpl.nasa.gov/horizons.cgi}} interpolated at the precise time-stamps of our measurements.
This allowed us to obtain unblurred maps providing a local (helioprojective) coordinate system having its origin at the Sun centroid (Solar-X and Solar-Y, \citealp{Thompson06a}).
FITS images are then produced and ready for scientific analysis (e.g. image RMS/sensitivity, dynamic range, and brightness profiles) through specific packages, such as {\sc SAOImage}\footnote{\url{https://sites.google.com/cfa.harvard.edu/saoimageds9}}, {\sc SunPy}\footnote{\url{https://sunpy.org}} and the Common Astronomy Software Applications ({\sc CASA})\footnote{\url{https://casa.nrao.edu/}}.
A selection of typical resulting images of the solar disk is represented in Fig.~\ref{fig:overview_images}.
\begin{figure}
\includegraphics[width=0.9\textwidth]{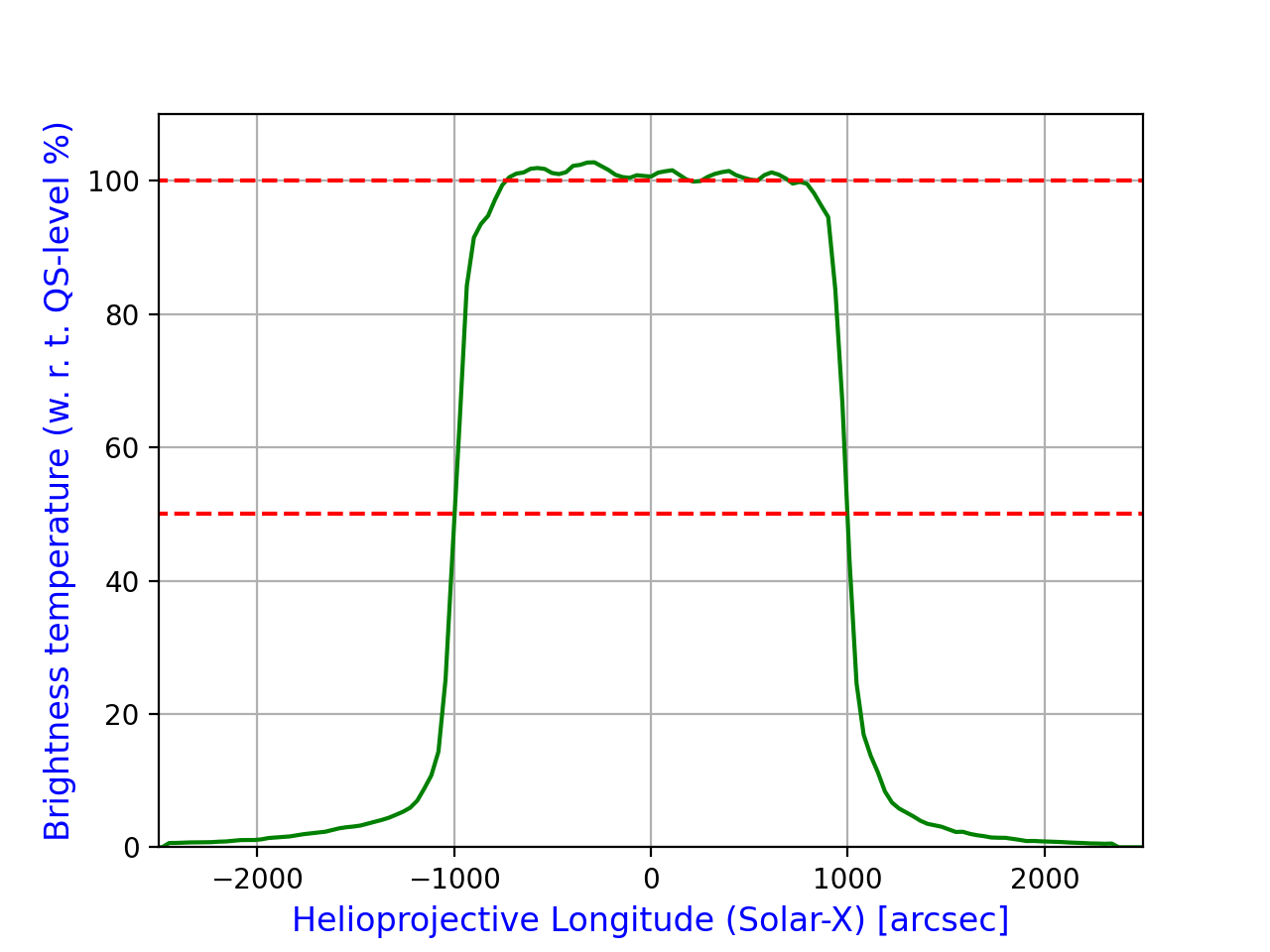}\centering
\caption{Example of a normalized brightness profile of a map sweep through the solar equatorial plane with Medicina radio telescope at $25.8$~GHz (3-Jan-2021).
The red dashed lines indicate the $100\%$ and $50\%$ QS-levels, respectively.
As described in \citet{Selhorst19}, the solar limb is conventionally defined at $50\%$ QS-level. The brightness fluctuations visible along the disk have a physical origin, while the broad and low-brightness wings outside the solar disk plateau are mostly due to the instrument beam pattern; the noise fluctuations are typically below $\sim$0.1\% level and they are not appreciable in the plot.
}
\label{fig:sun_radius}
\end{figure}
\begin{figure}
\centering
{\includegraphics[width=58mm]{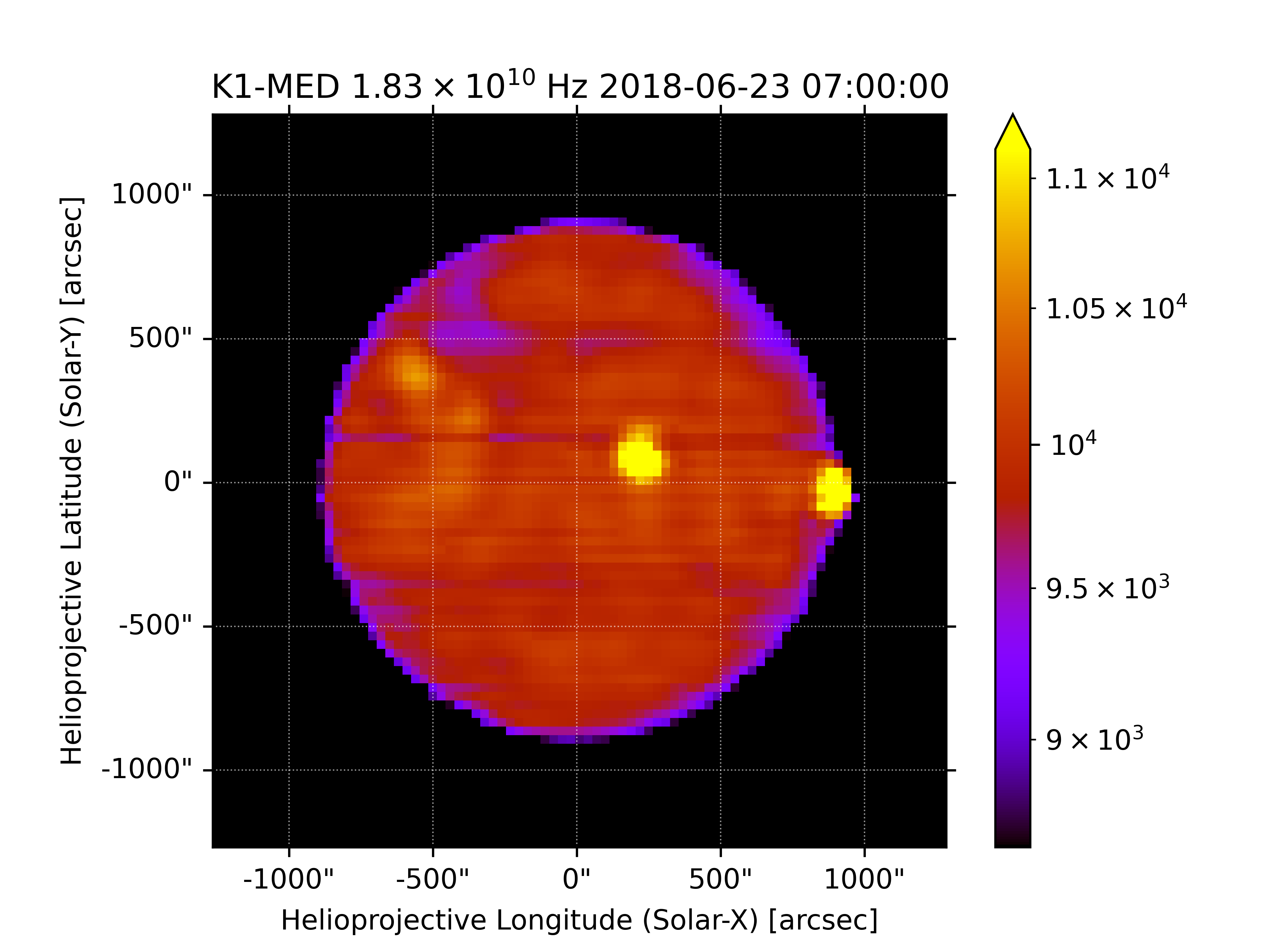}} \quad
{\includegraphics[width=58mm]{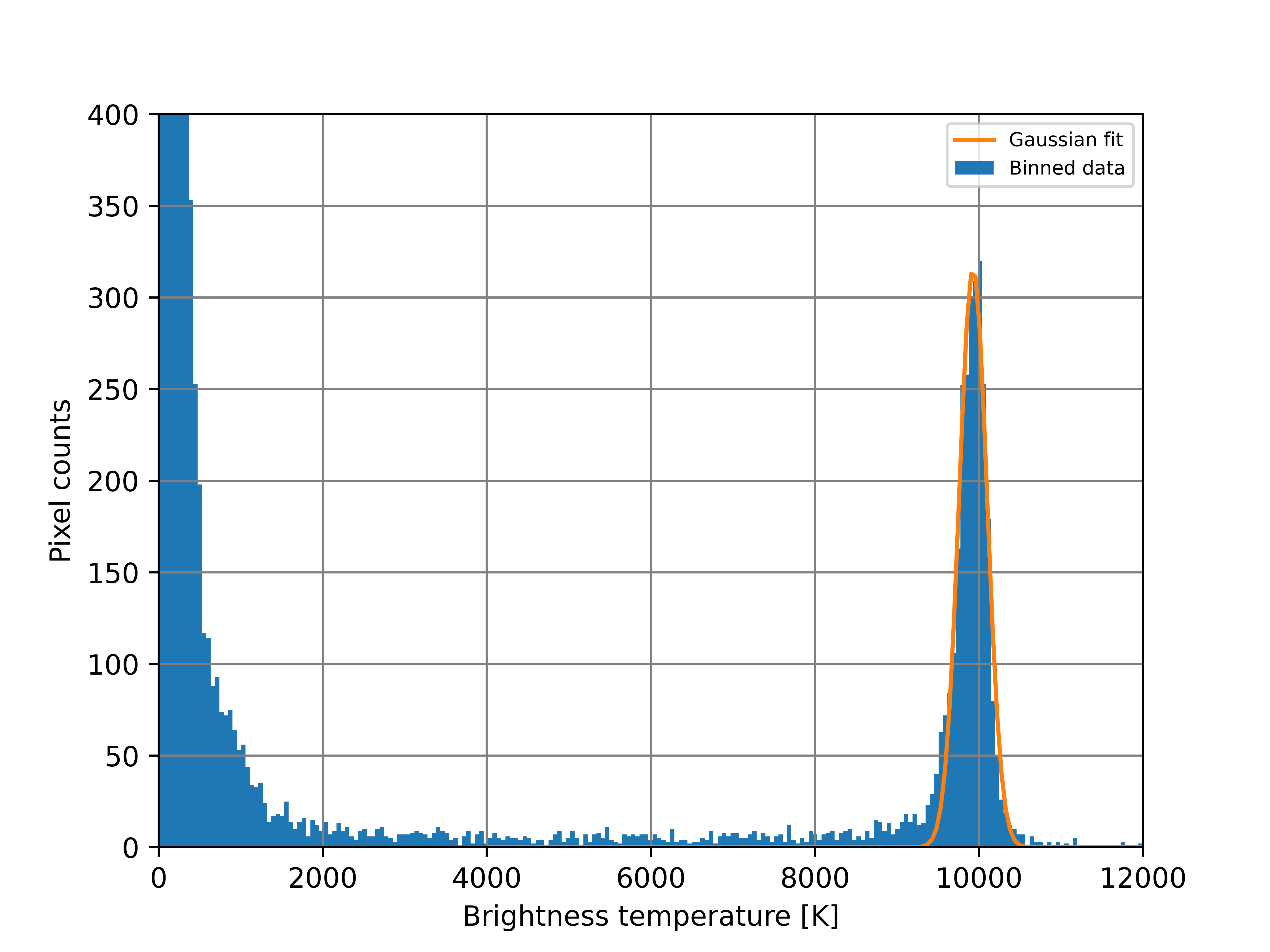}} \\
\caption{\textbf{(left)} Total intensity map of the solar disk at 18.3~GHz (bandwidth 1.2~GHz, and pixel size 0.6') obtained with the Medicina Radio Telescope on 23-Jun-2018.
The ARs SPoCA\,21867 (left, Footnote~\ref{foot:spoca}), NOAA\,12715 (SPoCA\,21859, center), and NOAA\,12713 (SPoCA\,21840, right) are evident in the image.
\textbf{(right)} Histogram of brightness distribution among pixels.
The Gaussian-like shape (orange line) in the histogram corresponds to the QS brightness distribution.
The low-brightness tail of the distribution is due to the fading brightness gradient of the area outside the solar disk near the limb.
}
\label{fig:histogram_maps}
\end{figure}

An example of typical brightness profiles of our maps is shown in Fig.~\ref{fig:sun_radius}, while an example of pixel brightness distribution, together with the corresponding solar map displayed with {\sc SunPy} tools, is given in Fig.~\ref{fig:histogram_maps}.

\subsection{Image calibration}
\label{subsec:im_calib}

In order to convert the raw maps (where intensity is expressed in "machine counts", directly proportional to the input signal level) to brightness temperature images, calibration observations are required.
As described in Sect.~\ref{sec:obs_setup}, the instrumental setup for solar observations involves significant additional signal attenuation with respect to non-solar radio telescope operations.
For this reason the amplitude of the standard calibration sources \citep{Perley17} is typically low, and a different calibration procedure is needed in order to obtain accurate measurements.

A first possibility is to compare the average raw counts in the image from the QS with literature data and published brightness information, in order to find a count-to-Kelvin conversion factor (self-calibration).
For this purpose a Gaussian fit of the image histogram (counts distribution among pixels) provides an accurate estimate of the average counts from the QS (see Fig.~\ref{fig:histogram_maps}).
It is worth noting that a Johnson's $S_U$-distribution \citep{Johnson49a,Johnson49b} provides a better fit of the QS brightness distribution with respect to the Gaussian fit in some cases, although this latter approximation does not significantly affect the calibration process.
For simplicity, we adopted the Gaussian fit in our data processing pipeline for the entire data set.

A brightness reference for our frequency range can be obtained from \cite{Landi08}.
The observed spectrum of the QS is characterized by a break at about 10~GHz, which corresponds to a brightness temperature of about 12,000~K referred to the center of the solar disk.
Above this frequency value, the spectrum is fitting ($\chi^2 = 0.032$) a logarithmic linear relation between the brightness ($T_b$, in units of Kelvin) and the frequency ($\nu$, in units of Hz): $log_{10} (T_b) = a + b \times log_{10}(\nu)$, where $a = 6.43$ and $b = -0.236$.
\cite{Landi08} do not provide a fit error estimate, but the measurements used in their model are typically affected by $\sim$ 5\% errors in the 10--20~GHz range and a much larger measurements spread is present above 30~GHz.
It is worth noting that no QS measurements were reported in the 18--30~GHz range in \cite{Landi08}.

A second calibration option can provide absolute brightness calibration also for the QS in K-band,
exploiting the relatively high dynamic range of SRT observations. 
The young and bright Cas~A (3C461) Supernova Remnant is a suitable flux calibrator that can be easily detected despite the application of the high attenuation levels needed for the observation of the solar disk.
Because of its high flux (about 2,400~Jy at 1~GHz), Cas~A is a common calibrator for high-frequency devices, and is circumpolar at the INAF radio telescopes latitudes.
However, since it is resolved both by Medicina and SRT in K-band (size about 5'), a different calibration approach is needed with respect to the OTF cross-scans on point-like standard calibrators.
OTF maps of Cas~A can be provided using the same observing parameters (such as frequency, and signal attenuation levels) adopted for the solar disk.
The counts-to-Kelvin conversion factor can be obtained from the comparison between the extrapolated flux of Cas~A at our observing frequencies (see Table~\ref{tab:calib_SRT}) and the source signal in our Cas~A image, accordingly to the procedure below.

The digital output (machine counts) of the instrument back-end is proportional to the source flux within the instrument beam.
The total flux of Cas~A ($S_{Jy}^{CA}$) can be compared with the total source counts ($C_{tot}^{CA}$) in the Cas~A raw image in order to estimate the calibration factor $f_{Jy}$ for counts-to-Jansky conversion:
\begin{equation}
\sum_i C_i^{CA} \Omega_{pix}^{CA} f_{Jy} = C_{tot}^{CA} \Omega_{pix}^{CA} f_{Jy} = S_{Jy}^{CA} 
\end{equation}
where $C_i^{CA}$ are the source counts for each pixel $i$ in the Cas~A image, and $\Omega_{pix}^{CA}$ is the pixel solid angle.
The corresponding brightness temperature calibration factor $f_{K}$ expressed in Kelvin in given by:
\begin{equation}
f_{K} = f_{Jy} \frac{c^2}{2 k_B \nu^2} = \frac{S_{Jy}^{CA} c^2} {2 k_B \nu^2 C_{tot}^{CA}  \Omega_{pix}^{CA}}
\label{eq:fk}
\end{equation}
where $\nu$ is the observing frequency, $c$ the light speed, and $k_B$ the Boltzmann's constant. 
The QS temperature is then obtained as $T_{QS} = f_K C_{QS}^{peak}$, where $C_{QS}^{peak}$ is the signal corresponding to the peak of the Gaussian (see Fig.~\ref{fig:histogram_maps}) that fit the QS emission distribution in the solar image (obtained during the same observing session and instrument setup used for Cas~A map).

Thus, the QS brightness temperatures obtained at different frequencies (see Sect.~\ref{subsec:qs_fluxes} and Table~\ref{tab:calib_SRT}) can be applied to generic raw solar images in order to obtain fully calibrated brightness maps through $T_{pix} = T_{QS}/C_{QS}^{peak} \times C_{pix} = f_K C_{pix}$, where $T_{pix}$ is the brightness temperature corresponding to the raw signal $C_{pix}$ in a given pixel.

In summary, a Gaussian fit (orange line, Fig.~\ref{fig:histogram_maps}) of the histogram section corresponding to the solar disk provided an estimate of the average instrumental signal from the QS (the Gaussian peak), scaled to Kelvin units through the calibration process.
The related Gaussian width $\sigma_{disk}$ corresponds to the standard deviation value of the QS brightness temperature as reported in the Appendix (Table~\ref{T:obs_summ}).

\section{Data analysis}
\label{par:obs_summ}

During the first three years of SunDish project, that included early development and observing test stages (early 2018 -- December 2020), we acquired and analyzed 169 solar maps with Medicina and SRT (see the Appendix, Table~\ref{T:obs_summ}). 
The resulting solar images are available for inspection and download in the web site dedicated to the project\footnote{\url{https://sites.google.com/inaf.it/sundish}.}.
Our image archive is structured to ease multi-wavelength data exploitation and it will be regularly updated online a few hours after each new observation. 

In the first release of the image catalog reported here, about 20\% of the solar maps were significantly affected by RFI contamination, instrumental anomalies and/or by bad weather conditions: poor or variable sky opacity during the observations (e.g. $\tau \gtrsim 0.1$ and/or fluctuation $\Delta \tau \gtrsim 0.01$ at 18~GHz).
Despite these images are still suitable for a rough estimate of AR parameters and fluxes, they typically display significant artificial brightness gradients in the QS component that prevent accurate calibration and affect the measurement of $\sigma_{disk}$ (see the Appendix, Table~\ref{T:obs_summ}).
On the other hand, about 80\% of our images sample ("high/medium quality" entries in the archive) provided typical sensitivities of a few Kelvin, well below the observed physical solar disk brightness fluctuations ($>$60~K), and then were suitable for accurate calibration and robust brightness and spectral analysis.

\subsection{Quiet-Sun brightness measurements}
\label{subsec:qs_fluxes}

The application of the calibration factor $f_K$ (Eq.~\ref{eq:fk}) to the back-end counts, obtained from the peak of a Gaussian fit of the QS histogram (see Fig.~\ref{fig:histogram_maps}), provides an estimate of the QS brightness temperature at the observed frequencies. This fit is not significantly affected by ARs contributions near solar cycle minimum.
We applied this procedure to selected SRT observations with optimal weather conditions both for the calibrator Cas~A and the solar mapping (five high-quality SRT maps were used in the process).
The Cas~A flux was extrapolated to our observing epochs and frequencies from \cite{Vinyaikin14}.
We provided an accurate error analysis considering the propagation effects of different uncertainties in the calibration process: Cas~A flux errors (including those arising for secular variation of the source flux), radio background contributions and sky opacity fluctuations (Cas~A and the Sun are typically observed at different elevations).
The overall calibration errors depend on frequency and range between 1.5 and 3~\%.
The resulting QS average brightness at the considered observed frequencies is shown in Table~\ref{tab:calib_SRT} together with the assumed flux values for Cas~A and their errors.
Our QS measurements are compatible with the model of \citet{Landi08} within a few percent discrepancies.
However, we note that both our measurements at 24.7 and 25.5 GHz are slightly higher ($>$1$\sigma$ level) than those extrapolated from the above literature reference.
These QS measurements will be further refined by the integration of more data and the improvement of Cas~A flux estimations, and presented in a future dedicated paper.
\begin{table}
\caption{QS brightness levels obtained from the image calibration process using high-quality SRT data. $\nu_{obs}$ is the central observing frequency; \textit{$Cas~A_f$} lists the SNR Cassiopeia~A integrated fluxes (and related observation epochs); \textit{$T_{QS}$} is the measured QS brightness temperature; \textit{$Fit_{dev}$} expresses the percentage deviation from the expected value extrapolated from \cite{Landi08}.}
\label{tab:calib_SRT}
\begin{tabular}{cccc}   
\hline                  
$\nu_{obs}$  & $Cas~A_f$                   & $T_{QS}$   & $Fit_{dev}$  \\
{[GHz]}      & [Jy]                        & [K]             & [\%]    \\
\hline
18.8         & 247.9 $\pm$ 5.7 (Oct--2020) & 10099 $\pm$ 154 & 0.24    \\
24.7         & 205.3 $\pm$ 4.8 (Oct--2019) & 9799 $\pm$ 268  & 3.24    \\
25.5         & 201.1 $\pm$ 4.7 (May--2019) & 9764 $\pm$ 223  & 3.65    \\
\hline
\end{tabular}
\end{table}

The sensitivity of our solar images is calculated through the measurement of the background brightness temperature fluctuations in map regions far from the solar disk emission.
Image sensitivities (RMS) are typically in the range $0.7$--$8.5$~K for Medicina and $0.1$--$0.3$~K for SRT.
Due to disk bleeding into the broad wings that slightly affect also the edge of the maps, these RMS estimations represent an upper limit of the actual image sensitivity. On the other hand, instrumental noise on the very bright solar disk is typically about an order of magnitude higher than the image sensitivity.

\begin{figure} 
\centering
\includegraphics[width=1.0\textwidth]{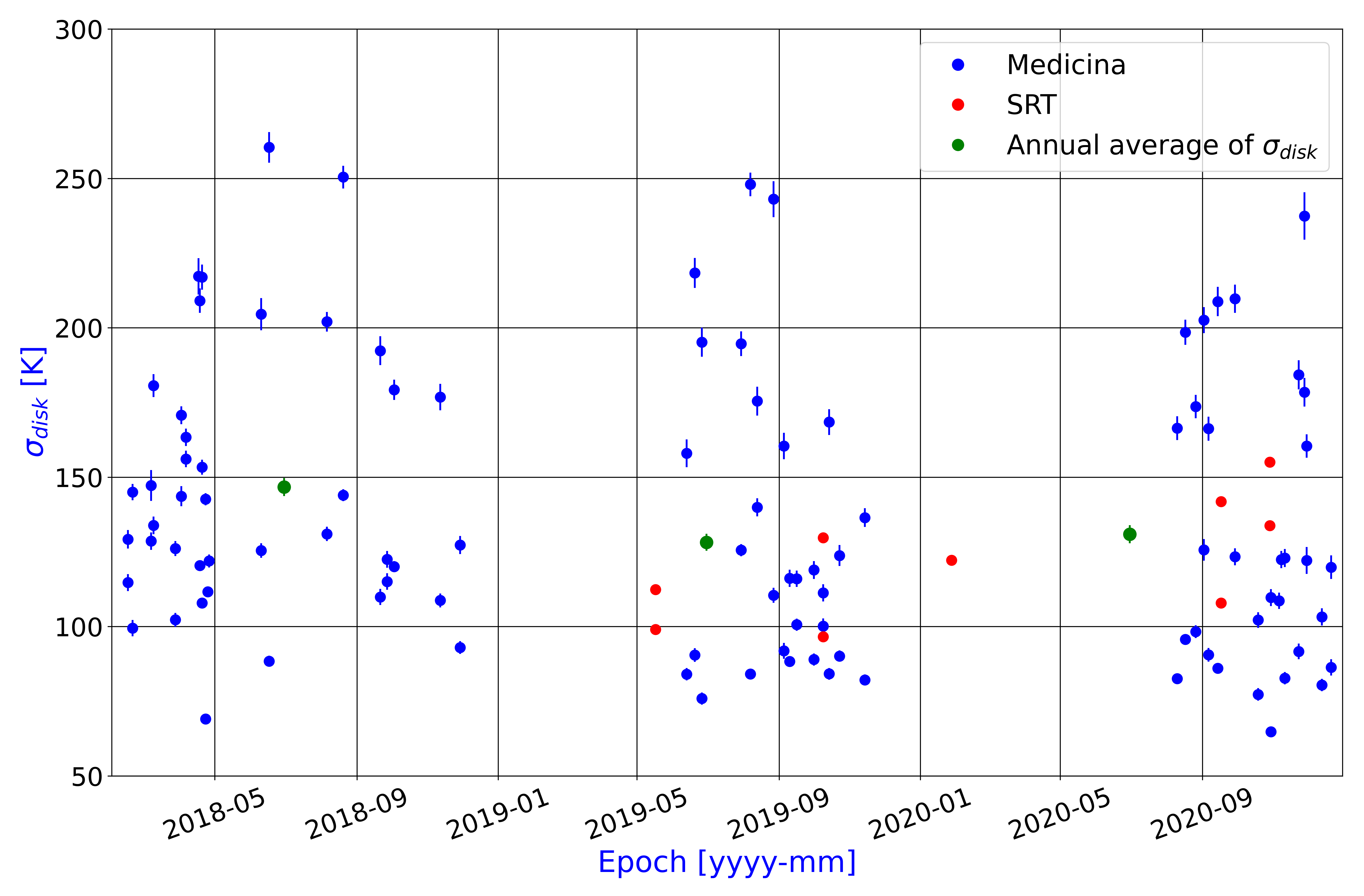}\centering
\caption{Temporal evolution of the width values $\sigma_{disk}$ of the Gaussian-like shape in the histogram of brightness distribution among pixels showed in Fig.~\ref{fig:histogram_maps} (orange line).
$\sigma_{disk}$ represents the standard deviation of the solar disk brightness distribution with respect to the QS level reported in Table~\ref{tab:calib_SRT} (only measurements related to medium/high-quality images are reported in the plot).
Green circles indicates the annual average of $\sigma_{disk}$.
}
\label{fig:plot_temp_sigma}
\end{figure}

The fit of QS brightness distribution provides $\sigma_{disk}$ values (see the Appendix, Table~\ref{T:obs_summ}) higher ($>60$~K) than the typical image sensitivity.
These fluctuations are mostly of physical origin and possibly related to the solar activity level.
In particular, the analysis of the local radiometric information allow us to estimate the apparent weather-related brightness temperature fluctuations on the solar disk. 
During typical observations in good/optimal weather conditions (high-quality maps) and for a typical observing period of 3~hours, these fluctuations are $< 6$~K at 18.8~GHz and $< 18$~K at 25.5~GHz.
The worst case during the observations can sporadically reach apparent weather-related brightness temperature fluctuations of about 100~K, significantly overlapping with actual physical fluctuations.
In these cases the effect of sky opacity fluctuations on the image is evident due to striping features (related to the OTF scanning procedures) and very different from the morphology of actual physical fluctuations on the disk (unrelated to the scanning pattern).
A temporal evolution of $\sigma_{disk}$ is plotted in Fig.~\ref{fig:plot_temp_sigma}.
The analysis of the possible variability of this parameter during the solar cycle requires long-term monitoring.

A preliminary estimate of the mean equatorial solar radius as defined in Fig.~\ref{fig:sun_radius} is $R_{eq} \sim 980$~arcsec at $25.8$~GHz,
in agreement with observations at 37~GHz performed by MRO \citep{Selhorst19}.
A detailed analysis of the solar radius, limb brightness and their possible evolution during the solar cycle (e.g., \citealp{Costa99,Emilio2000,Kuhn04,Selhorst05,Selhorst08,Selhorst11,Menezes17,Kosovichev18,Selhorst19}) is beyond the scope of the present investigation and will require further measurements.
We verified through convolution procedures on simple disk models with the beam pattern that the very broad wings in the profile in Fig.~\ref{fig:sun_radius} can be accounted for without invoking significant emission outside the solar disk plateau, as expected from optically thin free-free radio emission in these regions. These broad wings are mostly produced by secondary lobes of the beam pattern. Accurate beam pattern modeling and imaging deconvolution procedures will be discussed in future works.

\subsection{Active Regions fluxes and spectra}
\label{subsec:ar_fluxes}

Spots of enhanced radio emission due to ARs are present in most ($\sim 80\%$) of our 2018--2020 data set, even around the epoch of solar cycle minimum.
The identification of the ARs and the definition of their parameters -- e.g. the calculation of their fluxes and spectra -- is a non-trivial task.
It requires specific tools for pattern recognition and well-defined conventions in order to proficiently share data for multi-wavelength exploitation together with other very different solar imaging instruments.
We performed these operations through the implementation of a dedicated code developed in Python, called {\sc SUNDARA} (SUNDish Active Region Analyser, \citealp{Marongiu21}), adapted for our purposes from \cite{Marongiu20}.
It is based on several Python packages, such as {\sc Photutils}\footnote{\url{https://photutils.readthedocs.io/en/stable/}}, usually adopted for detecting and performing photometry of astronomical sources \citep{Bradley19}, and {\sc SunPy}, an open-source package for solar data analysis \citep{Mumford20}.
{\sc SUNDARA} receives as input the solar images (in standard FITS format) provided by the data processing pipeline described in Sect.~\ref{sec:data_an}, and in a few minute processing time it provides the identification of ARs and the estimation of their brightness temperatures, fluxes and spectral indices.

The {\sc SUNDARA} procedure unearths AR candidates rising from the QS brightness level through different algorithms that search for patterns consistent with an elliptical 2D-Gaussian kernel (see \citealp{Marongiu21} for details).
AR candidates are short-listed accordingly to a brightness threshold of $2\sigma_{disk}$ (see the Appendix,  Table~\ref{T:obs_summ})
of the QS level; this threshold was proven to avoid fake AR detection, providing enough sensitivity for the identification of known AR that are already included in public archives, and that typically exceeds the QS brightness level in our band of T$_{ex} > 1,000$~K $> \sigma_{disk}$.

Each AR candidate is then modelled through an elliptical 2D-Gaussian with noise, as described in \cite{Marongiu20}, where the free parameters are the AR helioprojective coordinates (Solar-X and Solar-Y; \citealp{Thompson06a}), the amplitude (peak brightness), the size (semi-axes and rotation angle of the ellipse), and the background noise (related to QS level local fluctuations).

\begin{figure} 
\centering
{\includegraphics[width=59mm]{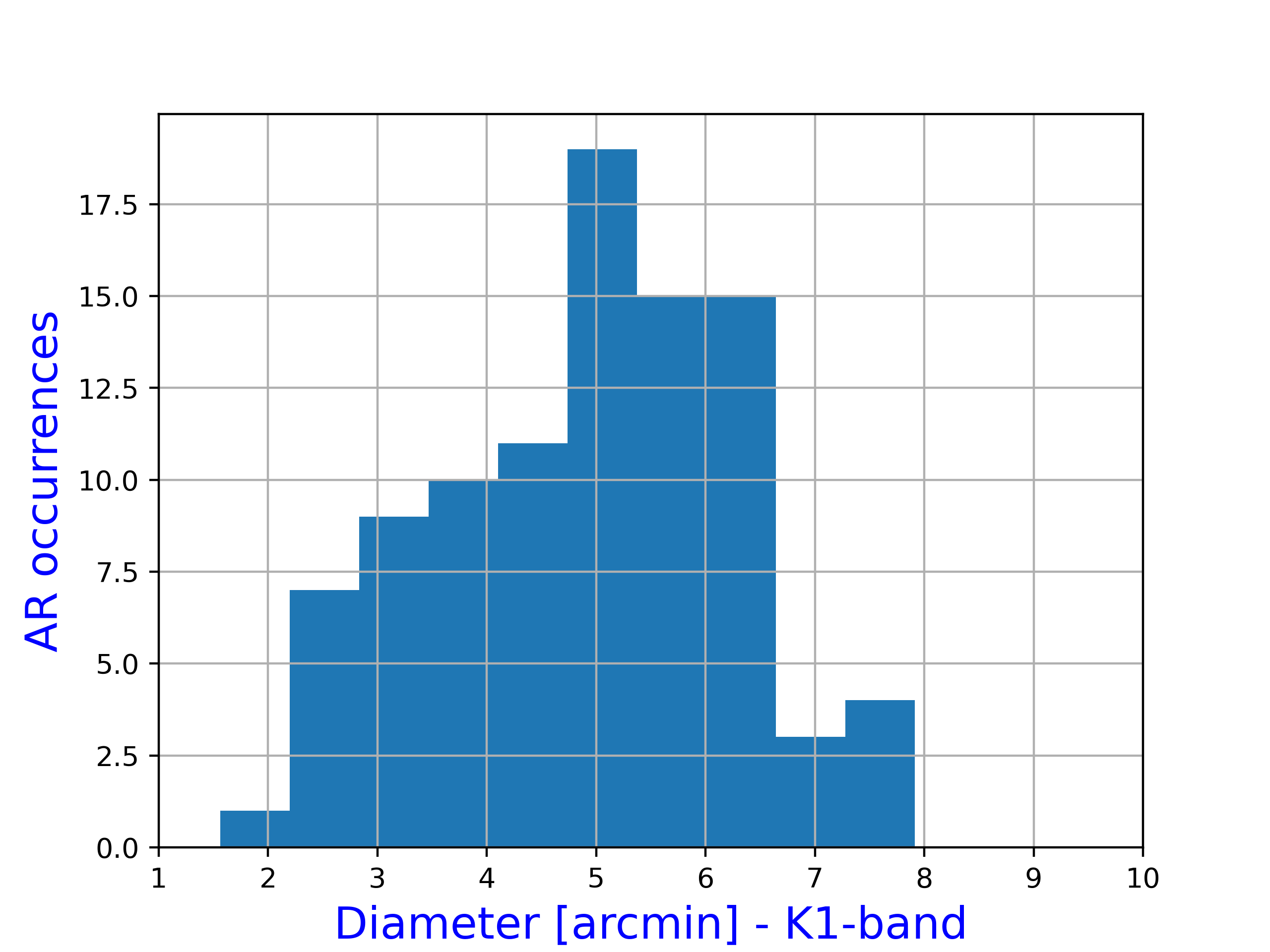}}\quad
{\includegraphics[width=59mm]{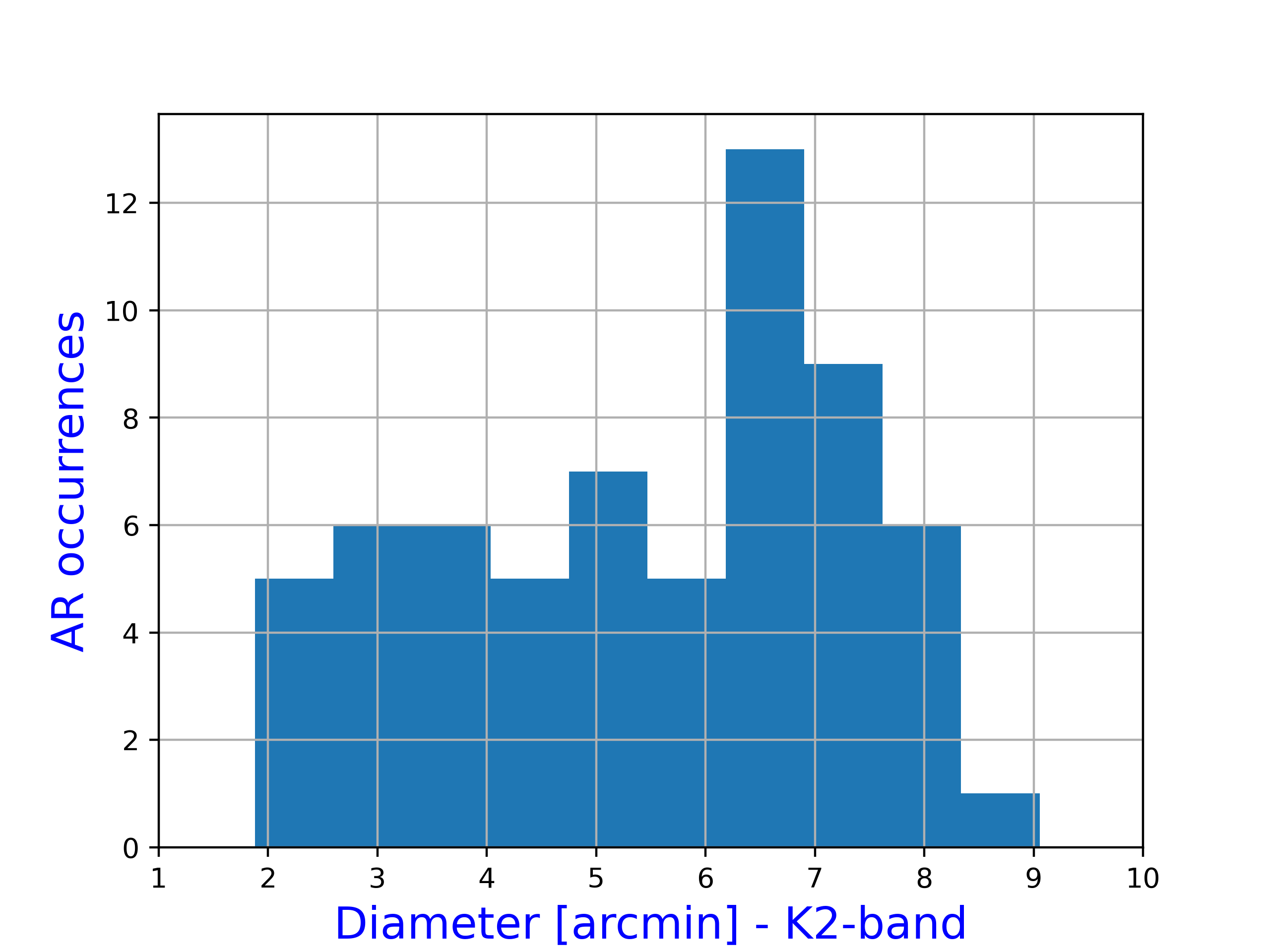}} \\
\caption{
Histogram of AR mean diameters in the 18--23~GHz (K1) and 23--26~GHz (K2) frequency ranges.
}
\label{fig:hist_diameter}
\end{figure}

Considering that usually the size and location of ARs detected in radio domain have a good coincidence with those observed in the high-frequency part of the electromagnetic spectrum (e.g., \citealp{Efanov72,Righini77,Silva05}), ARs are then associated or identified in terms of spatial position with known features (if any) provided by the multi-wavelength standard catalog Heliophysics Event Knowledgebase (HEK, \citealp{Hurlburt12})\footnote{\url{https://www.lmsal.com/hek/index.html}}.
These results applied to our first solar image catalog are included in "ar$\_$id" column of Tables~\ref{Tab:fluxes} and \ref{Tab:spec_index} in the Appendix.
To compare AR sizes with their parameters typically described in the literature (e.g. \citealp{Silva05}), we conventionally take the mean diameter (expressed as FWHM, in units of arcmin) obtained through the sum of semi-axes of the elliptical 2D-Gaussian fit.
A histogram of the AR mean diameters is shown in Fig.~\ref{fig:hist_diameter} for the 18--23~GHz (K1) and 23--26~GHz (K2) frequency ranges; these diameters are approximately the same at both radio bands as expected from our data resolution, and their values range from $1.6$ to $9.1$~arcmin, with a mean and standard deviation of $4.9 \pm 1.4$~arcmin at K1-band, and $5.5 \pm 1.8$~arcmin at K2-band, respectively.
The average size and standard deviation of our ARs\footnote{{\sc SUNDARA} calculates the AR area considering the modeled 2-$\sigma$ standard deviation along the semimajor/semiminor axes of the 2D-Gaussian function for each AR candidate.} are $30.4 \pm 16.2$~arcmin$^2$ for K1-band, and $37.8 \pm 23.0$~arcmin$^2$ for K2-band.
These AR mean diameters are compatible with those obtained from other radio/mm observations (\citealp{Silva05,ValleSilva20}) with NoRH ($17$~GHz), ALMA ($107$ and $238$~GHz, \citealp{Wootten_alma09}), and the Solar Submillimeter Telescope (SST, $212$ and $405$~GHz, \citealp{Kaufmann08}).
Thus, ARs are typically resolved in our images.

\begin{figure} 
\centering
{\includegraphics[width=59mm]{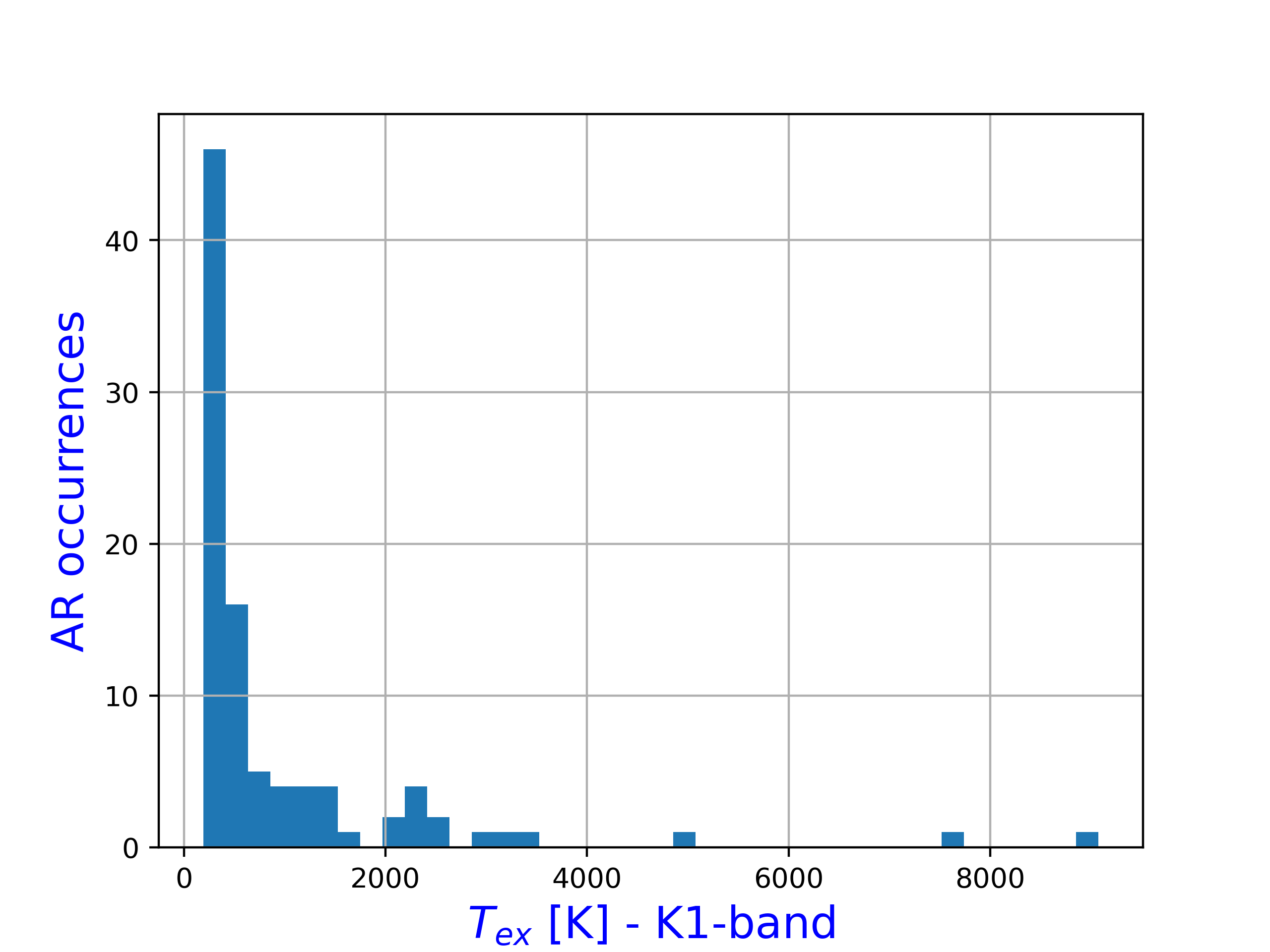}}\quad
{\includegraphics[width=59mm]{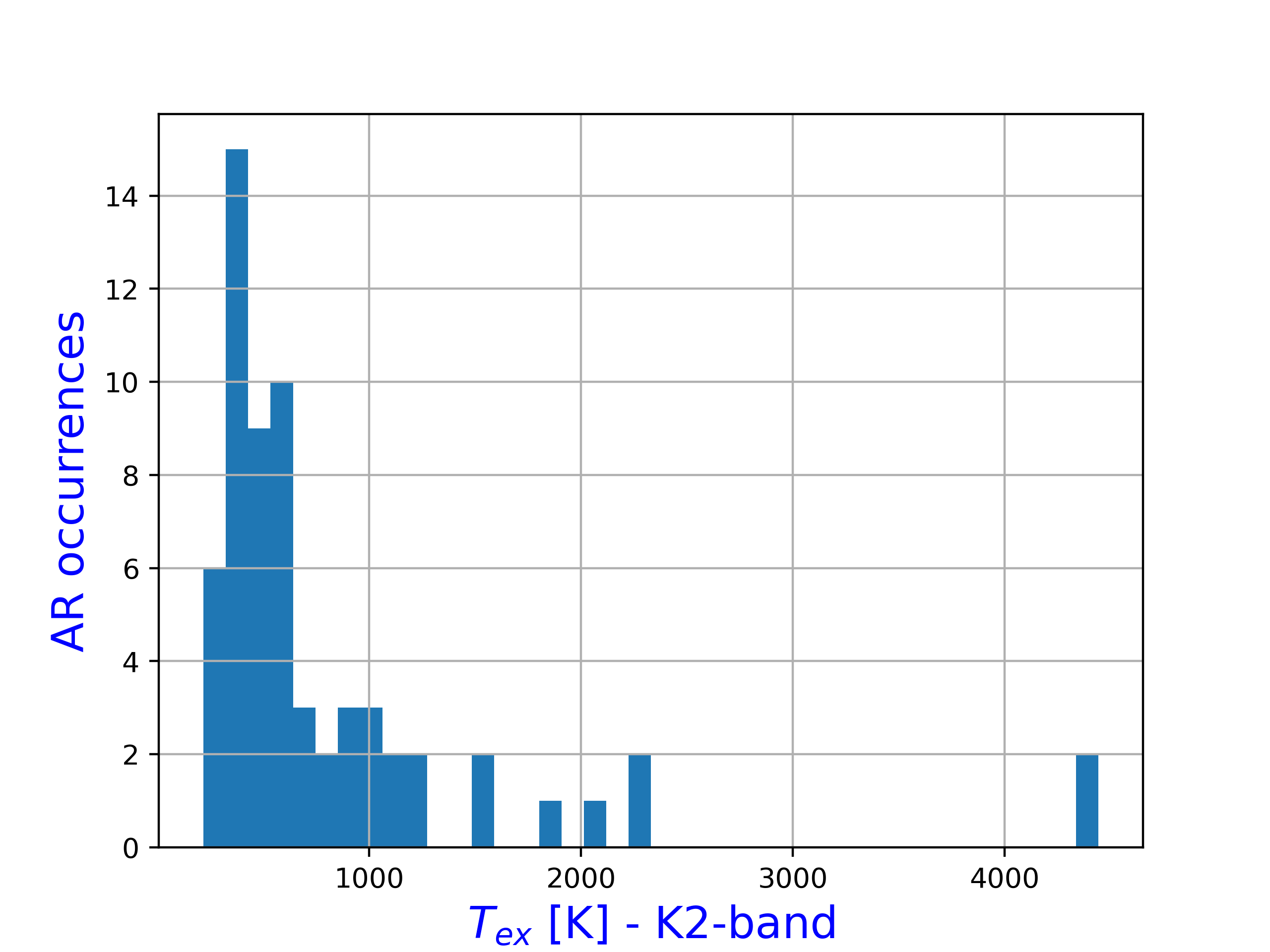}} \\
\caption{Histogram of the AR occurrences with excess brightness temperature $T_{ex}$ above QS values in K1-band (18--23~GHz, left panel) and K2-band (23--26~GHz, right panel).
}
\label{fig:hist_t_ex}
\end{figure}

The excess brightness temperature of ARs above QS levels, $T_{ex}$, is trivially defined as $T_b - T_{b(QS)}$, where $T_b$ and $T_{b(QS)}$ are the maximum brightness temperature of the AR and the QS temperature, respectively.
The error on $T_b$ is provided by the calibration uncertainties described in Sect.~\ref{subsec:qs_fluxes} ($\sim$ 2.5\% error on average, depending on frequency).
Statistical errors are typically of the order of $\sim$ 0.1\% due to very high signal-to-noise-ratio ($>10^4$) of solar disk brightness (typical image sensitivity $< 10$~K).
The obtained $T_{ex}$ distributions are shown in Fig.~\ref{fig:hist_t_ex}.
At both K1 and K2 bands, $T_{ex}$ values range between $\sim$~200 and over 2000~K, with median values of 430 and 560~K at K1 and K2-band, respectively.
AR emissions in our 2018--2020 data set exceed the QS brightness level up to a factor $\sim$ 2, although the average $T_{ex}$ is only $\sim$ 5\% of QS intensity.

The AR total fluxes $S_{\nu(AR)}$ (sfu units) are extracted from an elliptical region doubling the semi-axis of the 2D-Gaussian fit. They are calculated through the integration of the AR brightness inside that region.
Together with their uncertainties, they are given by:
\begin{equation}
S_{\nu(AR)} = 10^{22} \frac{2 k_B \nu^2}{c^2} \Omega_{pix} \sum_{pix} T_{b} \; [sfu],
\label{EQ:AR_flux_tot}
\end{equation}

\begin{equation}
\sigma_{S_{\nu(AR)}} = S_{\nu(AR)} \sqrt{(\sigma_{f_{cal}}/f_{cal})^2+(\sigma_{T_{b}}^{stat}/<T_{b}>)^2/N_{pix}} \sim \frac{S_{\nu(AR)} \sigma_{f_{cal}}}{f_{cal}}  \; [sfu],
\label{EQ:AR_flux_tot_err}
\end{equation}

\noindent where $N_{pix}$ the number of pixels (with solid angle $\Omega_{pix}$) associated to the AR region, $\sigma_{f_{cal}}/f_{cal}$ the calibration fractional error, and $\sigma_{T_{b}}^{stat}$ the statistical error corresponding to the image RMS.

In Table~\ref{Tab:fluxes} we provided ARs peak brightness and fluxes (in units of sfu) both in terms of total emission (including the QS component contribution) and as spatially integrated brightness excesses with respect to the QS flux level; the AR flag "C" in the "Notes" column indicates ARs located inside a confused region\footnote{This flag is connected with the quality of the background (the solar disk) around the extraction region (AR), in terms of the percentage variation $P_v$ between the fluxes obtained through the extraction region corresponding to $n = 2$ and $n = 5$ times the fitted semi-axes level; in particular, C flag appears when $P_v \geq 30\%$.\label{footnote:conf_reg}}.

For the observations where AR brightness/flux information (including lower and upper limits) is available at two or three frequencies ($\sim 80$ solar maps, characterized by high/medium quality), we provided spectral index ($S_\nu \sim \nu^{\alpha}$) measurements for $\sim 90$ ARs (see the Appendix, Table~\ref{Tab:spec_index}).
We performed the spectral index analysis of all the detected ARs comparing their peak brightness temperature with and without  the inclusion of the QS component (see Figs.~\ref{fig:istog_alfa_t_peak} and \ref{fig:istog_alfa_t_excess}, respectively); we also provided flux densities spectral index values (Fig.~\ref{fig:istog_alfa_tot}).
For the calculation of these latter values, we used the same extraction area for the fluxes estimated at different frequencies.
As extraction area, we considered the minimum AR size detected among the same ARs identified at the same epoch (and radio telescope) at different observing frequencies; this criterion allows us to prevent possible biases due to overestimated AR sizes, possibly characterized by a complex morphology.

From the histograms of the spectral indices (Fig.~\ref{fig:istog_alfa_tot}) we note the presence of values significantly outlying the main distribution.
Most of them correspond to actual physical values after discarding those partially ascribable to data analysis biases as: (1) observing noise caused by weather conditions (such as clouds) near the AR that might have contaminated the analysis (such as in the case of the 25.8~GHz map on 1-Oct-2019, where there is SD-AR\,M00037 characterized by $\alpha_{T_{ex}} > 4.1$), (2) background contamination around ARs (e.g., the AR SD-AR\,M00051 observed on 19-Oct-2020 with $\alpha_{T_{ex}} > 4.3$, characterized by a high AR extraction region of $> 50$~arcmin$^2$), (3) AR detection on the edge of the solar disk (for example in the case of the AR NOAA\,12778 / SPoCA\,24770\footnote{Spatial Possibilistic Classification Algorithm (SPoCA) is a multi-channel unsupervised spatially-constrained fuzzy clustering algorithm \citep{Barra05,Delouille12,Verbeeck14}, developed to segment the Extreme Ultra Violet (EUV) images of the Sun into 3 categories (QS, AR and coronal holes).\label{foot:spoca}} observed on 30-Oct-2020 with $\alpha_{T_p} = 1.5$).

\section{Discussion}
\label{par:disc}

Despite INAF radio telescopes being facilities not dedicated to solar monitoring, we were able to perform weekly observations on average in this early science phase of the project, with good performances.
The sensitivities of our solar images in the 18--26~GHz frequency range are of a few Kelvin or less, with an instrumental noise and sky opacity fluctuations on the bright disk usually $<$10~K (Sect.~\ref{subsec:ar_fluxes}), about one order of magnitude lower than the typical standard deviation of the QS brightness distribution (see Table~\ref{T:obs_summ} and Fig.~\ref{fig:plot_temp_sigma}).

We accurately measured QS brightness levels through the adoption of the very bright SNR Cas~A as a flux calibrator (see Sect.~\ref{subsec:qs_fluxes}); in good observing conditions, the overall calibration errors do not exceed 3\%, and this accuracy can be improved with further calibration campaigns.
This robust image calibration procedure allows us to perform a systematic analysis of the radio parameters of ARs detected in most of our observing sessions.

{\sc SUNDARA} (Sect.~\ref{subsec:ar_fluxes}) has successfully identified brightness excesses regions in about 70\% of solar observing sessions characterized by maps with medium/high quality (see the Appendix, Tables~\ref{Tab:fluxes} and \ref{Tab:spec_index}).
Such structures included both known ARs and other radio unidentified areas (with unclear reference in the literature).
In order to measure the AR size and total excess flux we relied on our algorithms, and chose a specific convention \citep{Marongiu21}.
Since other observatories could employ different methods, a common convention and a data extraction method are crucial in order to perform a careful multi-frequency spectral analysis together, and hence to obtain coherent and standardized results.

\begin{figure}
\includegraphics[width=0.8\textwidth]{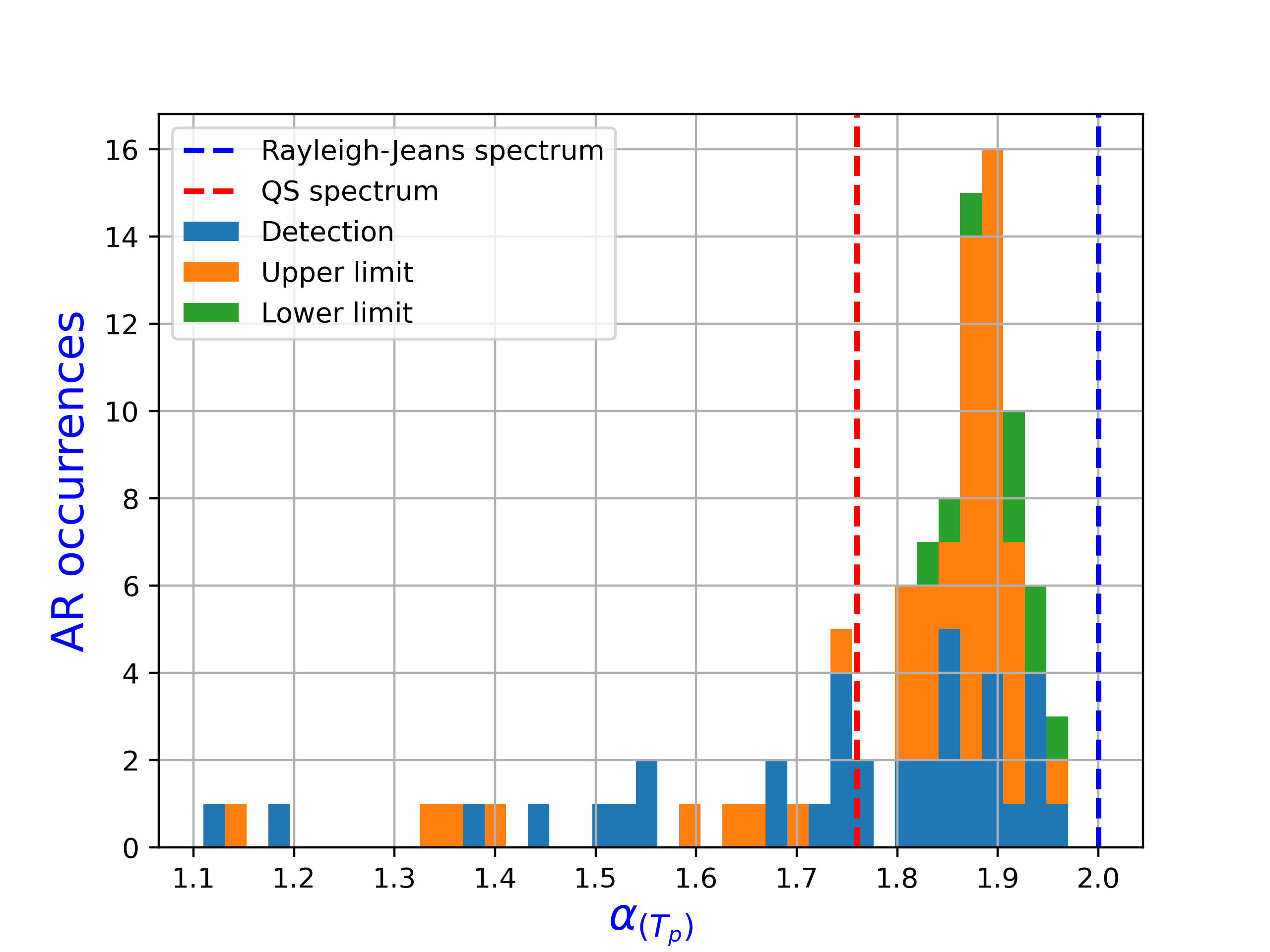}\centering
\caption{Histogram of the spectral index values calculated from the maximum brightness temperature $T_p$.
The data are binned in $40$ bins.
Blue counts indicate detections; orange and green counts show upper limits and lower limits, respectively. 
Dashed lines indicate the spectral indices of Rayleigh-Jeans (blue), and of the QS-level as estimated by \citet{Landi08} (confirmed by our QS measurements).
}
\label{fig:istog_alfa_t_peak}
\end{figure}

Comparing these AR features with those reported at other frequencies\footnote{EUV and optical data are obtained thanks to facilities on-board several missions, such as Hinode (\url{https://www.isas.jaxa.jp/home/solar/}), and Solar Dynamics Observatory (\url{https://sdo.gsfc.nasa.gov/})},
we note that AR details and substructures -- as the complex AR morphology \citep{McIntosh90} and magnetic structures \citep{Jaeggli16} -- are typically not resolved in our data (see Fig.~\ref{fig:overview_images}), while the bulk emission of many ARs is typically slightly larger than our spatial resolution (see radio beam width in Tables~\ref{tab:info_med} and \ref{tab:info_SRT}).
On the other hand, we typically detect rich network structures on larger scales, compatible with the "semi ARs" described in a recent work by \cite{Kallunki20}, based on single-dish radio observations at 37~GHz; they claimed that these "semi ARs" -- detected even during the minimum phase of the solar cycle -- can be indicative of persistent solar activity in the radio domain, suggesting that these weak radio brightenings are mostly related to coronal hole features and magnetic bright points.
The brightness excess temperatures with respect to the QS level ($T_{ex}$) are below $\sim$ 1,000~K in most cases (see Fig.~\ref{fig:hist_t_ex}); this suggests that in our data the chromospheric network shows typical temperature fluctuations of $<$ 10\%, except from sporadic ARs with total emission nearly doubling the QS brightness level. 

\begin{figure}
\includegraphics[width=0.8\textwidth]{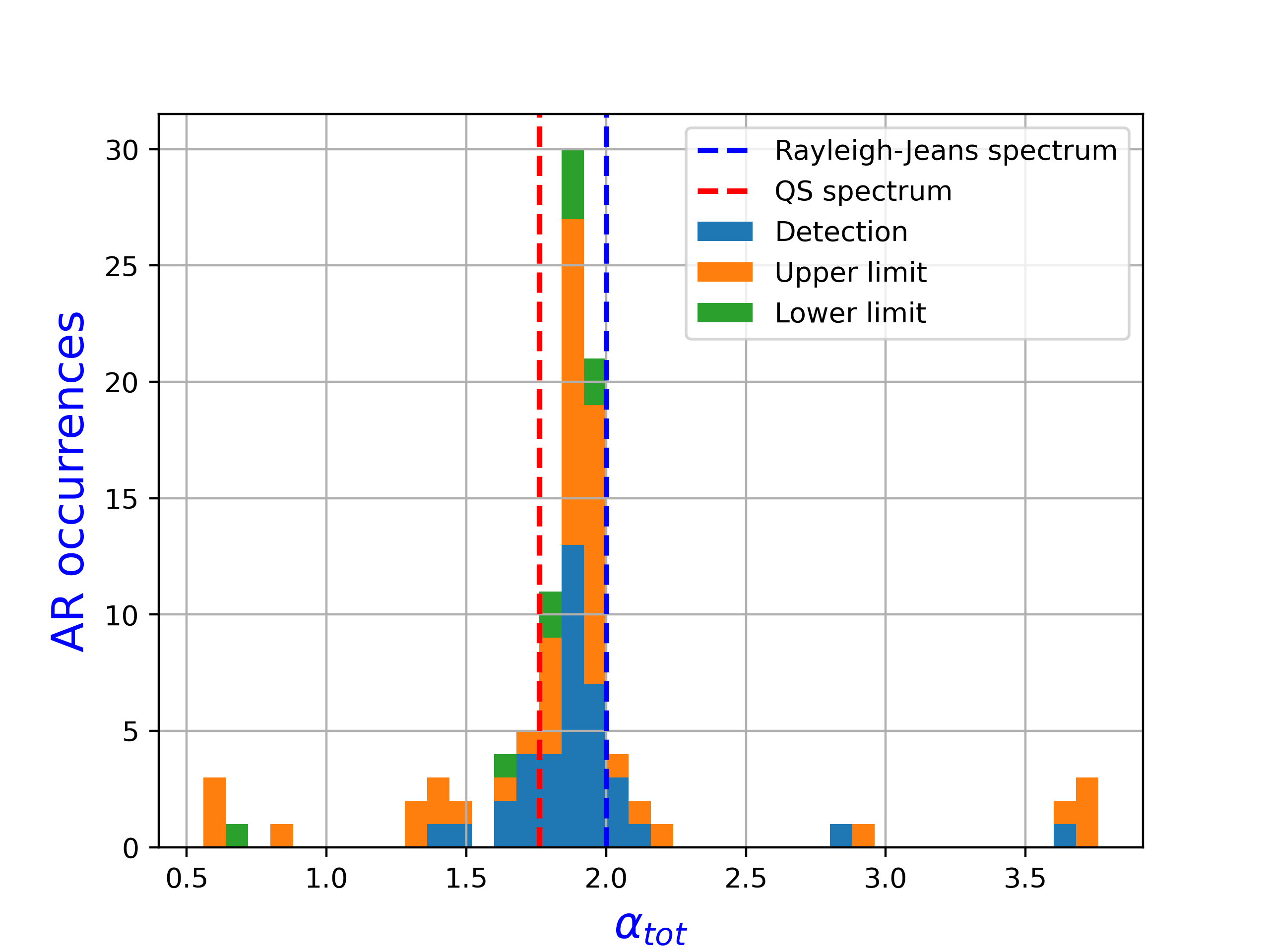}\centering
\caption{Histogram of the flux density spectral index values $\alpha_{tot}$.
See the caption of Fig.~\ref{fig:istog_alfa_t_peak} for a full description of the histogram.
}
\label{fig:istog_alfa_tot}
\end{figure}

The analysis of 169 solar maps in the frequency range 18--26~GHz (Table~\ref{Tab:spec_index}) provides important clues about AR spectra.
In Fig.~\ref{fig:istog_alfa_t_peak} we report the spectral index distribution obtained by comparing AR peak brightness temperatures (including the QS component) at different frequencies,
while in Fig.~\ref{fig:istog_alfa_tot} we plot the AR spectral distribution related to integrated total fluxes.

These distributions, peaked between the Rayleigh-Jeans (RJ) and the QS spectrum ($\alpha_{QS} \sim 1.75$ in our frequency range) as derived from \citet{Landi08}, suggests the presence of emission significantly deviating from simple thermal bremsstrahlung in the optically thick regime ($\alpha_{RJ} = 2$) as observed in ARs at higher frequencies \citep{Silva05,GimenezDeCastro20,ValleSilva20}.
We note that our distribution in  Fig.~\ref{fig:istog_alfa_t_peak} peaks around $\alpha_{T_p}$$\sim$$1.9$;  
optically-thin spectral index from thermal bremsstrahlung is generally regarded as closer to this value than to 2.0 due to the logarithmic dependence of the Gaunt factor on frequency.

The fact that many of the detected spectral indices in Fig.~\ref{fig:istog_alfa_t_peak} are higher ($\alpha_{T_p} > 1.8$) than QS values could suggest a thermal gradient within the vertical structure of typical ARs that differs from QS emission models.
Furthermore, in addition to thermal bremsstrahlung, the concurring presence of sporadic gyro-magnetic components in the AR emission could contribute to spectral softening.
In fact, gyro-resonance emission has been proven to peak up to $\sim 20$~GHz (see e.g. \citealp{Selhorst08}), although in most cases this component is observed at lower frequencies (2--5~GHz; \citealp{Kakinuma62,Nindos02}).
The presence of strong gyro-synchrotron emission is unlikely in our data due to the low chance of observing a flare during the scans, although small-scale energy release events (nanoflares) cannot be excluded.
More AR statistics is needed to precisely assess the actual distribution of unusually soft spectra, also considering the presence of many upper limit measurements.
\begin{figure} 
\includegraphics[width=0.8\textwidth]{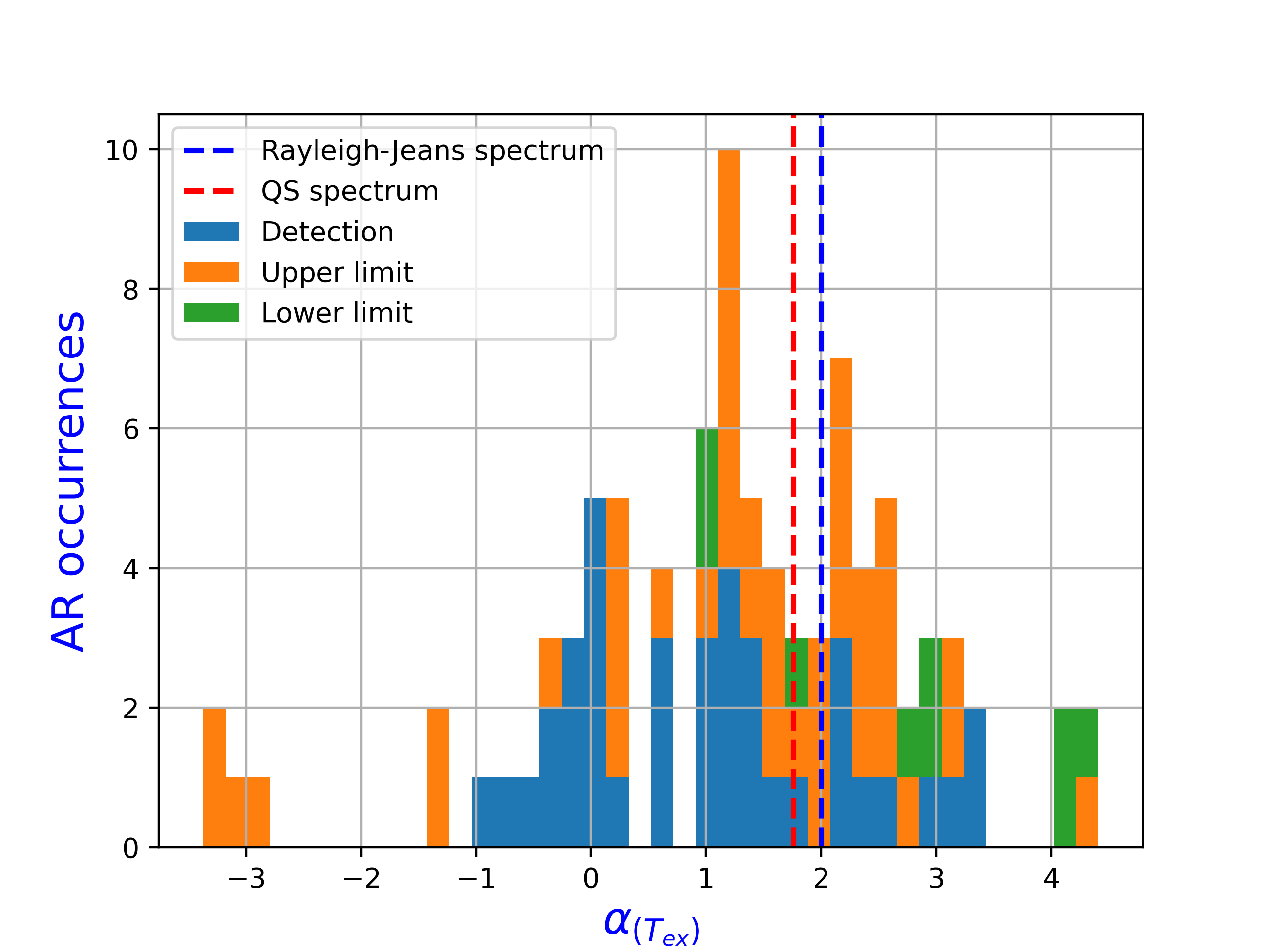}\centering
\caption{Histogram of the spectral index values calculated from the maximum brightness temperature excess $T_{ex}$.
See the caption of Fig.~\ref{fig:istog_alfa_t_peak} for a full description of the histogram.
}
\label{fig:istog_alfa_t_excess}
\end{figure}

The AR spectral index distribution in Fig.~\ref{fig:istog_alfa_t_excess}, based only on the weak brightness excesses with respect to QS component (obtained by subtracting the QS level itself) emphasizes the characteristics of ARs emission and their non-thermal components.
Fig.~\ref{fig:istog_alfa_t_excess} shows that a significant number of spectra are relatively flatter than the thermal bremsstrahlung emission, corroborating the possible presence of gyro-magnetic emission components.
We could explain these spectra of peculiar AR events considering the contribution of a steep high-frequency tail provided by gyro-resonance emission peaking slightly below our frequency range.
It is important to note that the measured spectral indices represent an average value within the beam and thus differ from the actual values rising from the smaller-scale unresolved regions that emerge from the diffuse free-free emission.
For example, an hypothetical steep tail of gyro-resonance emission with spectral index $\alpha=-7$ and contributing to $\sim$50\% of the total brightness at 18~GHz will result in an apparent flat spectral index when averaged with the free-free thermal emission covering the whole radio beam.

The gyro-resonance emission at radio frequencies has been studied by many authors in order to estimate the coronal magnetic fields, up to the 15--17~GHz range (e.g. \citealp{Akhmedov82,Alissandrakis93,Shibasaki94,White95,Nindos2000,Kundu01,Vourlidas06}).
They showed that the gyro-resonance emission for radio frequencies $\gtrsim 10$~GHz is produced in regions of magnetic fields with intensities of kG (located close to the transition region), and that at 17~GHz the lower limit for the magnetic field intensity to produce this emission (due to the third harmonic) is $\sim$ 2000~G at the photospheric level; moreover, the enhancement of gyro-magnetic effects when leaving the minimum of solar activity could cause a significant increase of the AR average brightness up to $\sim 10^5$~K, and beyond.

In Fig.~\ref{fig:istog_alfa_t_excess} we also note a significant number of spectra harder than RJ emission.
This could be related to gyro-resonance emission peaking above our frequency range (rising part of the spectral component) and requiring very high magnetic fields $> 2000$~G and/or higher-harmonics components. 
Imaging observations above 30~GHz reported in the literature are not conclusive about the possible presence of gyro-resonance spectral contributions in addition to free-free emission \citep{Selhorst08}.
In theory, at these frequencies gyro-resonance emission characterized by a typical magnetic field intensity of $\sim 2000$~G (at the photospheric level) is allowed for high harmonics (6--7); typical emission in the gyro-frequency lower harmonics (2--4) is more difficult to observe, because uncommonly high magnetic field intensities ($\sim 3000-6000$~G) would be in principle required (e.g., \citealp{Selhorst08,Valio20,Nindos20}).

A possible physical context about these spectral outliers could be related to local eruptive events inside ARs -- including flares and coronal mass ejections (CMEs) -- crucial to investigate the AR dynamics.
A gyro-magnetic emission enhancement could be a precursor of flaring gyro-synchrotron emission from the region of anomalous energy release with a sub-relativistic plasma in moderate magnetic fields \citep{Vatrushin89}, and modelled according to several approaches (e.g., \citealp{Somov86,Podgorny12}).
Broadband analysis of these eruptive events shows changes -- a few days prior to the event -- in their microwave/radio spectra, suggesting both the emergence of a new AR photospheric magnetic field and shifting movements of the sunspots \citep{Shibasaki11,Bogod12,Borovik12,Abramov-maximov13}.
These changes can be used as a predictive criterion for eruptive (geoeffective) events on the Sun.

\begin{figure} 
\centering
{\includegraphics[width=59mm]{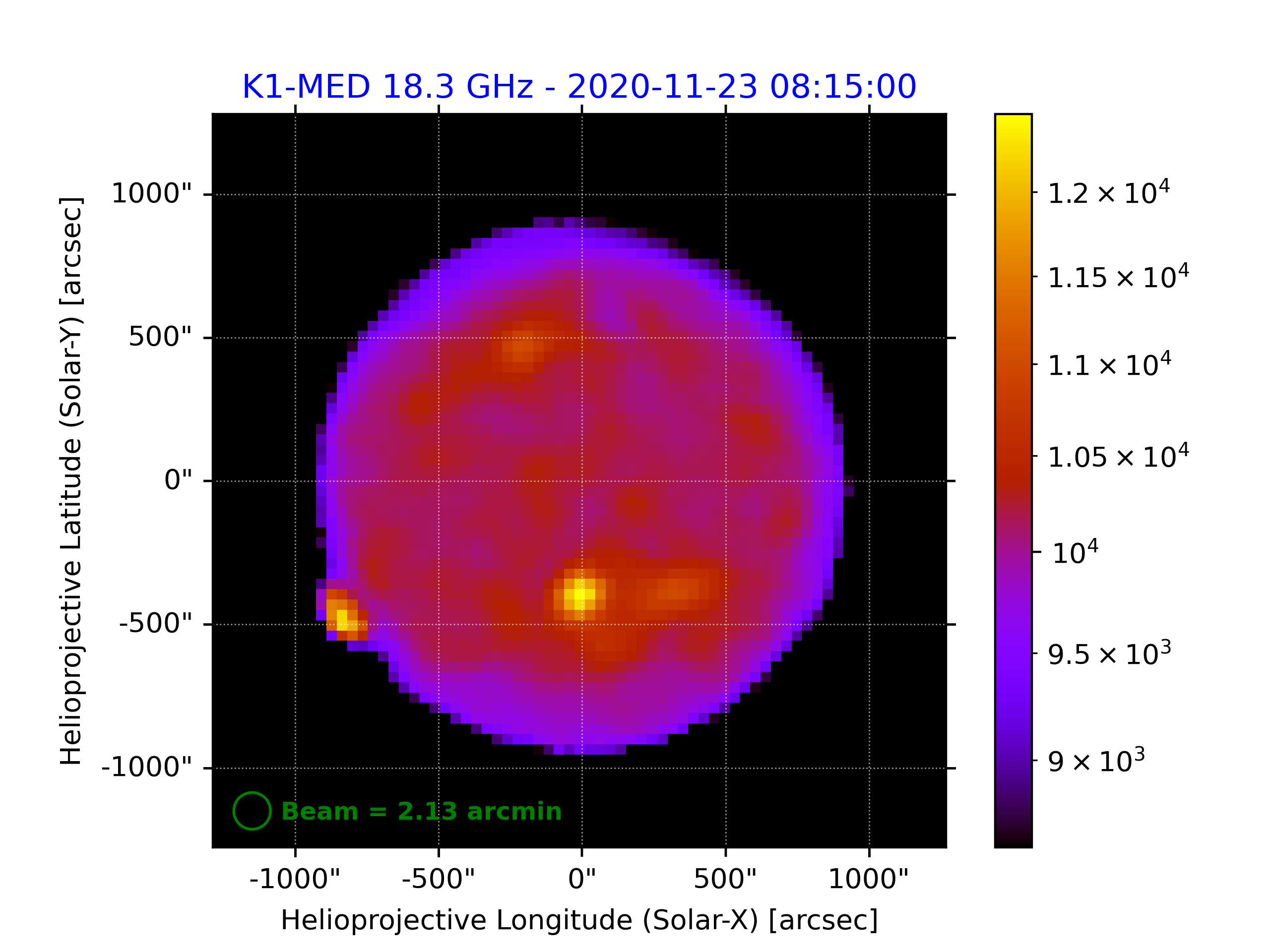}}\quad
{\includegraphics[width=59mm]{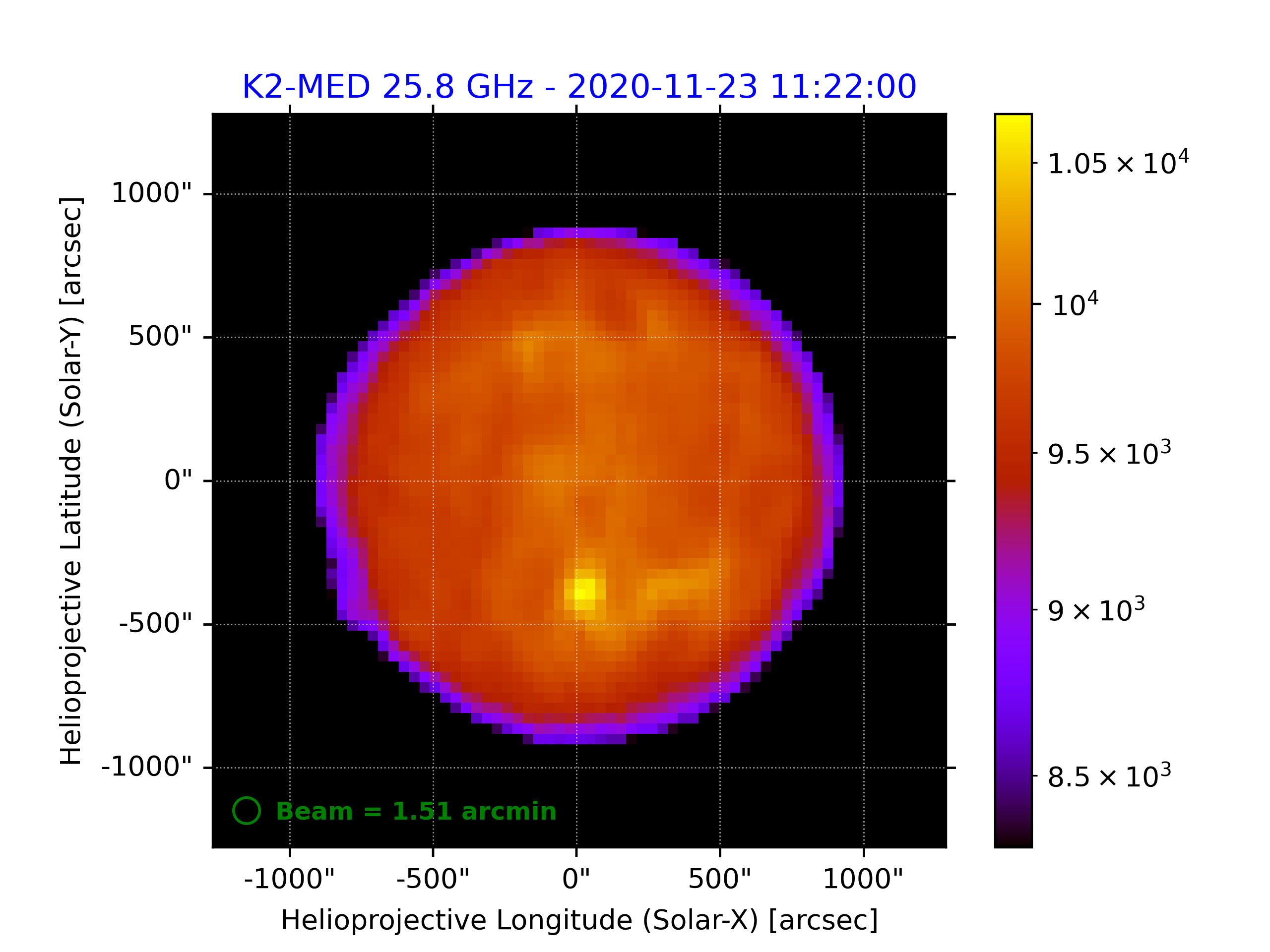}} \\
\caption{
Total intensity map (in units of Kelvin) of the solar disk at 18.3~GHz (left) and 25.8~GHz (right) obtained with the Medicina Radio Telescope on November 23$^{th}$ 2020.
AR NOAA\,12786 / SPoCA\,24827 is visible only at 18.3~GHz at the edge of the solar disk near the solar limb.
It represents a typical example of AR with very soft spectrum that can anticipate a subsequent C-class flare emission.
}
\label{fig:flare_november}
\end{figure}

Several papers (e.g., \citealp{Borovik12,Abramov-maximov13}) show a correlation between changes in the magnetic field configuration of the ARs and the eruptive events that have taken place one or two days prior in the same region.
This could lead to a change in the balance between the thermal and non-thermal emission component, with a prevalence of the gyro-magnetic emission that - for example - can produce an inversion from positive to negative values in the spectral index.
In fact, on November 23rd 2020, we observed a very soft spectrum ($\alpha_{T_P} < 1.1$ e $\alpha_{tot} < 0.6$) in AR NOAA\,12786 / SPoCA\,24827 and just a few hours later a C-class flare occurred arising from the same AR (Fig.~\ref{fig:flare_november}).
Extensive correlation studies with the eruptive events are needed in order to support this hypothesis.

It is worth noting that  anomalous spectral indices could be also related to sudden temporal variations of AR brightness between the non-simultaneous observation at different frequencies that are typically taken $\sim$ 1--2~hours from each other.
A recent work by \cite{Kazachenko20} shows that ARs observed in EUV band are characterized by temporal variability of irradiance of $\sim 3\%$ in a daily timescale.
Moreover, currently {\sc SUNDARA} is not able to distinguish two (or more) very close ARs with angular distance less than 2 beams of the receiver\footnote{For example, this is the case of SPoCA\,24770 / SPoCA\,24772 (on 29-Oct-2020 with $\alpha_{T_p} = 1.2$), SPoCA\,24827 / SPoCA\,24840 (on 28-Nov-2020 with $\alpha_{T_p} = 1.4$), and SPoCA\,24827 / SPoCA\,24850 / SPoCA\,24851 / SPoCA\,24853 (on 30-Nov-2020 with $\alpha_{T_p} = 1.5$), detected as a single AR.}.

Our results are compatible with other multi-frequency radio analysis of ARs (e.g., \citealp{Silva05,ValleSilva20}), where the contribution to $T_{ex}$ is at least partially attributable to gyro-resonance emission at frequencies $\lesssim 50$~GHz, while thermal bremsstrahlung is the only emission component seen at high frequencies (100--400~GHz).
\cite{Silva05} show that in the total flux density spectra ($\alpha_{tot}$) of ARs the cm-millimetric emission (17 and 34~GHz) exceeds systematically the expected fluxes from the optically thick free-free emission; this is due to the concurring presence of (1) gyro-resonance emission at 17~GHz \citep{Castelli66,Kaufmann68,Tapping01}, (2) optically thick thermal free-free emission from the upper chromosphere, and (3) optically thin thermal emission from the transition region and coronal sources.
A 3D solar atmospheric model, developed to reproduce the brightness temperature of a specific AR characterized by high degree of polarization ($85\%$) with radio observations at 17 and 34~GHz from NoRH \citep{Selhorst08}, showed that at 17~GHz the emission was successfully modeled as gyro-resonance, while at 34~GHz the emission was attributed to free-free radiation only.

In summary, the variety of AR spectra that we observed in the 18--26~GHz frequency range could be related to the variable contribution of gyro-resonance components that peaks in this range or in neighborhood frequencies.
Our findings seem to provide clues of gyro-resonance lines even slightly above 26~GHz, possibly associated with part of the $\sim$33\% of ARs that we detected with very peculiar hard spectral indices ($\alpha_{T_{ex}}>2$).

\section{Conclusions and prospects}
\label{sect:prosp}

We presented strategies, techniques and the scientific potentialities of the new solar observing system of INAF radio telescopes, with a focus on radio continuum imaging of AR features and related statistics.
Our solar monitoring aims to bridge the low-frequency radio observations -- mostly characterized by episodes of strong and variable gyro-magnetic emission -- with the chromospheric network depicted by thermal emission at higher frequencies (e.g., \citealp{Selhorst08}).
Our observations show that the so-far poorly explored 18--26~GHz frequency range harbors very dynamical processes in the chromosphere that require smart and frequent mapping of the entire solar disk.

We firstly focused on the assessment of the temperature brightness of the QS component and related calibration issues, exploiting the minimum of solar activity in the period 2018--2020.
Our early science studies on the calibrated solar disk images focused then on the identification of bright ARs and related spectral measurements.

The interesting variety of AR spectra in the 18--26~GHz will require simultaneous multi-wavelength analysis to disentangle all the emission parameters.
Such analysis should include neighborhood radio frequencies sharing data with other observatories as e.g. LOFAR and MRO: in general higher frequencies constrain free-free thermal emission, while lower frequencies provide information on the involved gyro-resonance emission (see e.g. \citealp{Nindos20}).
Despite these latter physical processes cannot be spatially resolved in our images, these studies could contribute to the assessment of the maximum observed frequency associated to gyro-resonance peak emission and its related magnetic field, when combined with polarimetric information.
Our AR spectral data offer clues about the possible existence of sporadic gyro-resonance emission above $\sim 26$~GHz even around the solar minimum.
Among future SRT/SDSA solar configurations, observing capabilities at new frequency bands will be implemented, simultaneously covering the X (8--9~GHz) and K (25.5--27.0~GHz) or X and Ka (31--33~GHz) bands that may play a crucial role to constrain the frequency range of AR gyro-resonance emission.
To date, this component has been detected till 15--17~GHz (see \citealp{Nindos20} and references therein), while at 34~GHz \citet{Selhorst08} modeled the emission as purely free-free.
Thermal components will be further constrained through the upgrading of SRT with the new receivers in Q-band (33--50~GHz) and W-band (75--116~GHz) whose operations are expected to start in 2023.
These instruments coupled with the new generation digital backend SKARAB (Square Kilometer Array Reconfigurable Application Board)\footnote{\url{http://www.tauceti.caltech.edu/casper-workshop-2017/slides/12_moschella.pdf}} will offer a wider dynamic range best suited to detect strong gyro-magnetic and synchrotron emission in perspective of the growing solar activity.

Polarimetric mapping of solar disk emission features could allow us to correlate magnetic information with AR spectra, helpful in disentangling gyro-magnetic contributions from free-free emission.
A very high degree of polarization ($>$ 30\%) is expected in sporadic ARs dominated by a  strong gyro-magnetic component \citep{Selhorst08}.
Full-Stokes polarimetric information is already available for SRT solar observations, while at Medicina station only dual polarization observing mode is presently available in the K-band.
The exploitation of such information will represent a next challenging milestone for this project, considering the expected low degree of polarization ($<$ 10\%, \citealp{Selhorst05}) of the QS component and most of disk features, and the need of dedicated polarimetric calibration procedure for solar observations.

AR radio observations will support empirical and physical models of spectral behavior that can anticipate an upcoming flare; the correlation between AR spectral index changes (due to incoming strong gyro-magnetic components from X to K bands) and subsequent flare occurrence in a few hours/days can be statistically investigated and it might represent a valuable forecasting probe for flare events contributing to the Space Weather warning/alert network. 

The additional complex network of diffuse and weaker emission structures typically present in our images  ("semi ARs" as defined by \citealp{Kallunki20}), as well as emission regions at lower temperature than the QS level possibly related to coronal holes, will require dedicated investigations.
The cross-comparison of our simultaneous 7-feed images\footnote{Accurate solar data calibration is presently available for the central feed only, despite image quality is good in most feeds.} provided by SRT observations clearly demonstrate the realm of such weak structures not related to instrumental gain fluctuations.

It will be interesting to monitor the evolution of the extension and intensity of ARs and the diffuse "semi AR" features during the progression of the solar cycle.
This monitoring can be supported by the study of secular evolution of the $\sigma_{disk}$ parameter (see Section \ref{subsec:qs_fluxes}, Fig. \ref{fig:plot_temp_sigma}), and its correlations with sunspots numbers (SSN) and other solar indices.
Several papers show the correlation between AR parameters and the solar cycle in the radio domain.
A recent work about spectral analysis with SST at $212$ and $405$~GHz during Cycles 23 and 24 \citep{GimenezDeCastro20} shows that the temporal evolution of AR brightness ($T_{ex}$) at these two frequencies follows the solar cycle, represented by the SSN.
A similar result was obtained by \cite{Selhorst14} at 17~GHz: a correlation between the AR brightness temperature and the solar cycle is attributed to the emergence of ARs characterized by strong magnetic fields ($\geq 2200$~G at the photosphere; \citealp{Vourlidas06}) and a gyro-resonance contribution, able to increase the AR brightness temperatures up to $10^5$~K (and beyond), if the 17~GHz gyro-resonance third harmonic ($\sim 2000$~G) occurs above the transition region \citep{Shibasaki94,Vourlidas06,Selhorst08,Shibasaki11}.

\begin{figure} 
\centering
{\includegraphics[width=59mm]{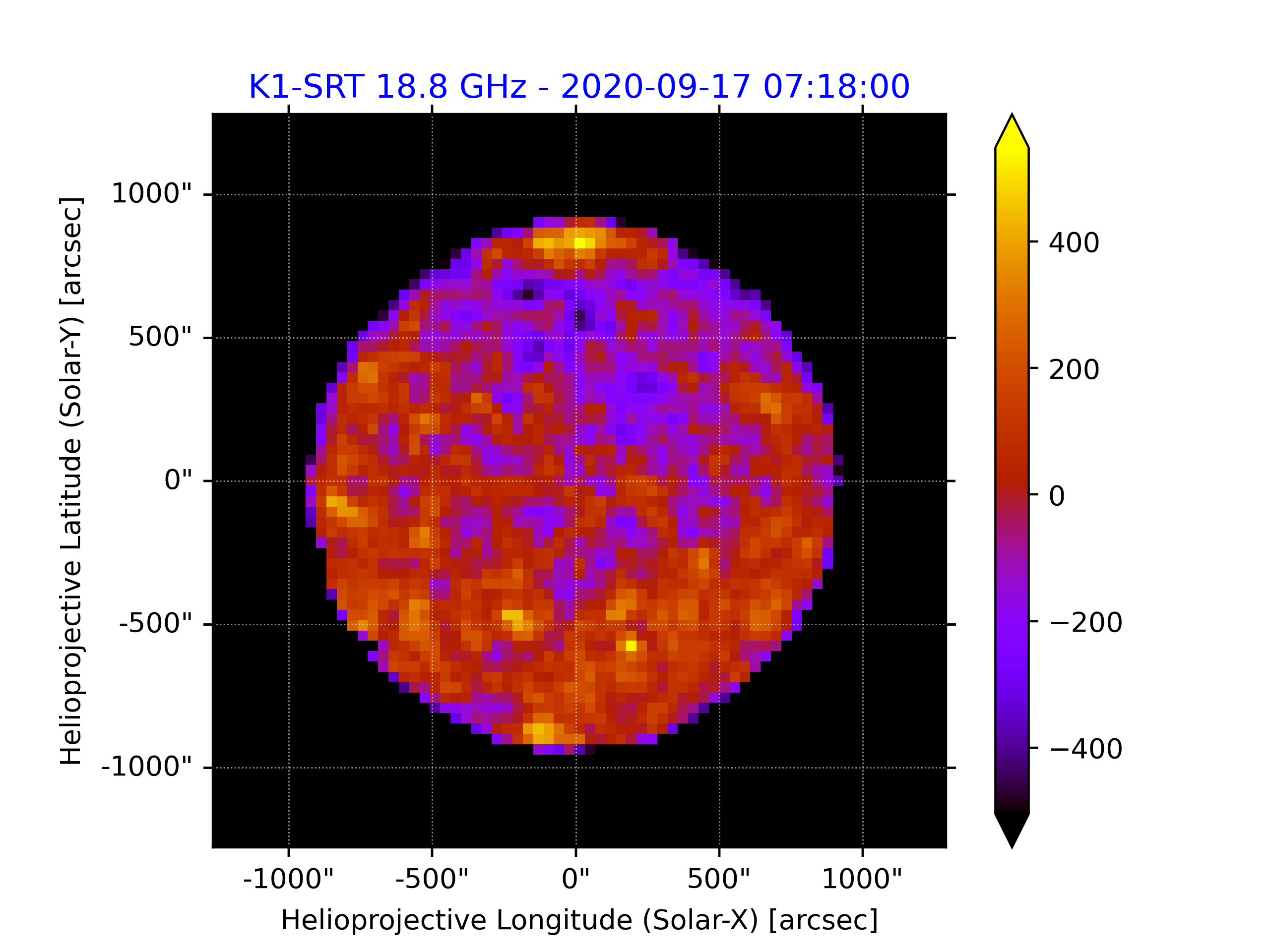}}\quad
{\includegraphics[width=59mm]{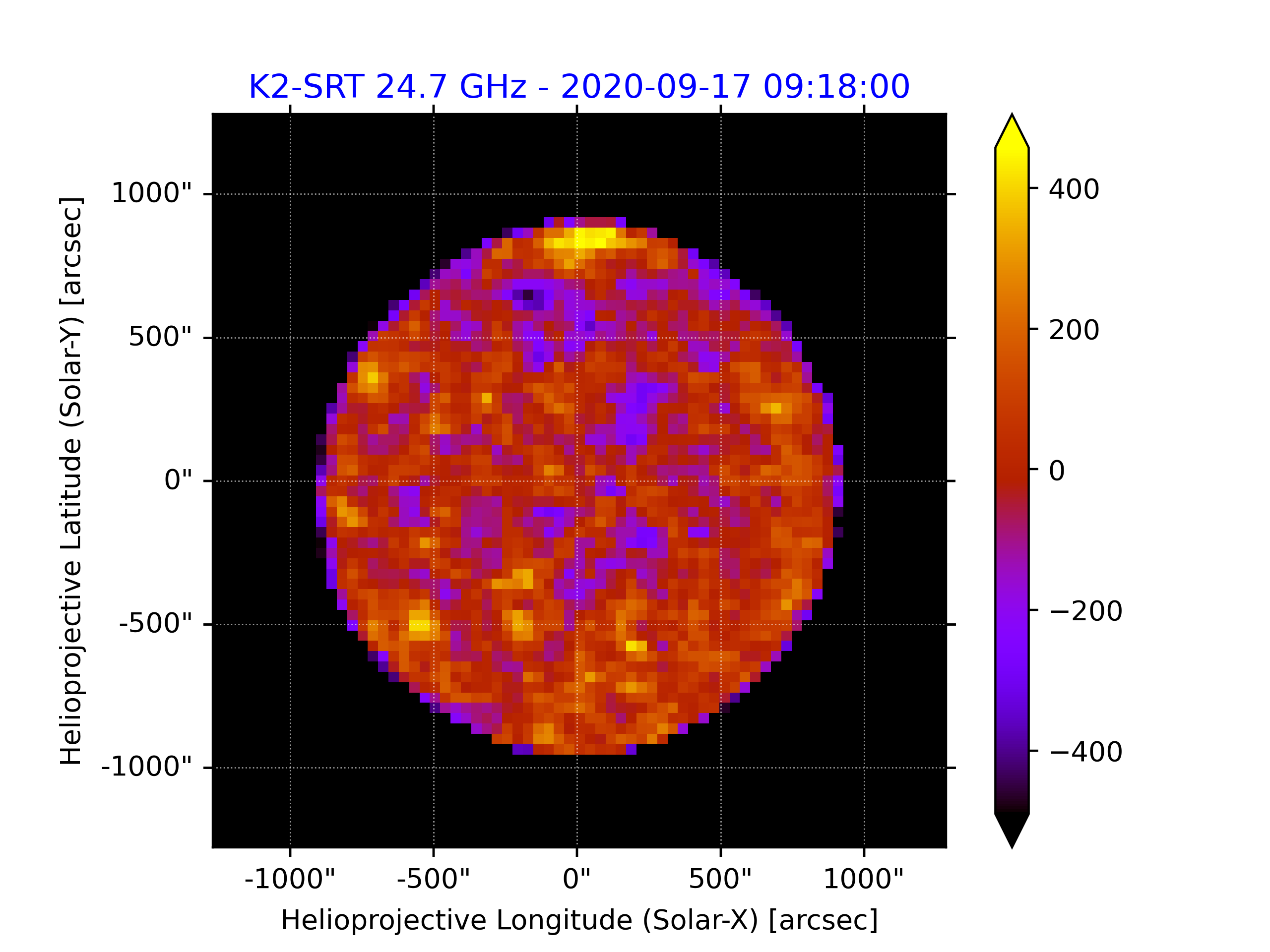}} \\
\caption{
Total intensity map ($T_{ex}$ brightness level) of the solar disk at 18.8~GHz (left) and 24.7~GHz (right) obtained with SRT on 17-Sep-2020, characterized by a typical episode of polar brightening.
}
\label{fig:ex_polar}
\end{figure}

The study of radio counterpart of other peculiar large-scale structures, such as coronal holes, loop systems, filaments, streamers and the coronal plateau, will also benefit from systematic long-term observations during the evolution of the solar cycle.
For example, interesting solar disk features as polar brightening (see. e.g. \citealp{Kosugi86,Shibasaki98}) are clearly detected in our data (e.g. on September 17$^{th}$ 2020 at SRT, Fig.~\ref{fig:ex_polar}).
The radio brightness in polar regions seems anti-correlated with solar activity as suggested by NoRH images at 17~GHz in the epoch range 1992--2013.
This effect could be related to the fact that around solar activity minimum -- characterized by a reduced presence of ARs -- polar regions are dominated by strong unipolar magnetic field that may enhance their brightness (see e.g. \citealp{Shibasaki11,Nitta14}).
For systematic studies of solar cycle dependence of polar brightening (as a probe of magnetic field variations during a solar cycle), long-term full-disk observations are required, and ideally starting from the present phase near minimum solar activity.

%
\begin{acks}

The Sardinia Radio Telescope is funded by the Italian Ministry of University and Research (MUR), Italian Space Agency (ASI), and the Autonomous Region of Sardinia (RAS).
The Medicina radio telescope is funded by the Italian Ministry of University and Research. Both radio telescopes are operated as National Facilities by the National Institute for Astrophysics (INAF). 

The Enhancement of the Sardinia Radio Telescope (SRT) for the study of the Universe at high radio frequencies is financially supported by the National Operative Program (Programma Operativo Nazionale - PON) of the Italian Ministry of University and Research "Research and Innovation 2014-2020", Notice D.D. 424 of 28/02/2018 for the granting of funding aimed at strengthening research infrastructures, in implementation of the Action II.1 – Project Proposals PIR01\_00010 and CIR01\_00010.

\noindent S.M. acknowledges contributions from the Italian Space Agency for ASI/Cagliari University grants no. 2019-13-HH.0 and no. 2020-34-HH.0.

\noindent This research used version 3.1.3 \citep{sunpy_community2020} of the SunPy open source software package \citep{Mumford20}. \\

\par

\underline{\textbf{This is a pre-print of an article accepted for publication in Solar Physics.}}

\end{acks}

\newpage
%
\appendix   

\begin{longtable}{l|ccccc}
\caption{Summary of Medicina and SRT observations.
\textit{ID} indicates the identification number for each single map, where the letters \textit{M} and \textit{S} specify the radiotelescope (Medicina or SRT); an asterisk flag is shown for images presenting significant artifacts.
\textit{Epoch} indicates the observation date, \textit{T} the acquisition time interval of the map, and $\nu_{obs}$ the central observing frequency; $\sigma_{disk}$ represents the standard deviation of the solar disk brightness distribution with respect to the QS level reported in Table~\ref{tab:calib_SRT}.
$AR_n$ indicates the number of identified ARs in each solar map.
}
\label{T:obs_summ} \\
\hline                   
{ID}    & Epoch      & T         & $\nu_{obs}$ & $\sigma_{disk}$    & $AR_n$ \\
        & [yy-mm-dd] & [UT]      & [GHz]       & [K]                &        \\
\hline \hline
\endfirsthead
\caption{Continued.} \\
\hline
{ID}    & Epoch      & T         & $\nu_{obs}$ & $\sigma_{disk}$    & $AR_n$ \\
        & [yy-mm-dd] & [UT]      & [GHz]       & [K]                &        \\
\hline \hline
\endhead
\hline
\endfoot
\hline
\endlastfoot
$M1$    & 18-02-15 & 11:20-12:20 & 24.1 & 129.8 $\pm$ 4.0   & 1 \\
$M2$    & 18-02-15 & 12:22-13:22 & 24.1 & 114.5 $\pm$ 3.6   & 1 \\
\hline
$M3$    & 18-02-19 & 10:14-11:30 & 24.1 & 102.4 $\pm$ 3.8   & 1 \\
$M4$    & 18-02-19 & 11:33-12:49 & 24.1 & 145.2 $\pm$ 3.5   & 0 \\
\hline
$M5$    & 18-02-27 & 09:39-10:50 & 23.6 & 160.2 $\pm$ 4.2   & 1 \\
$*M6$   & 18-02-27 & 10:52-12:02 & 23.6 & 269.4 $\pm$ 18.6  & 1 \\
\hline
$M7$    & 18-03-07 & 13:50-15:00 & 23.6 & 129.5 $\pm$ 3.9   & 1 \\
$M8$    & 18-03-07 & 15:02-16:05 & 23.6 & 148.0 $\pm$ 7.0   & 0 \\
\hline
$M9$    & 18-03-09 & 12:11-13:20 & 23.6 & 180.5 $\pm$ 5.1   & 1 \\
$M10$   & 18-03-09 & 13:22-14:33 & 23.6 & 134.0 $\pm$ 3.9   & 1 \\
\hline
$M11$   & 18-03-28 & 10:05-11:20 & 23.6 & 103.1 $\pm$ 3.1   & 0 \\
$M12$   & 18-03-28 & 11:22-12:37 & 23.6 & 126.3 $\pm$ 3.4   & 0 \\
\hline
$*M13$  & 18-03-30 & 09:45-11:00 & 23.6 & 137.1 $\pm$ 4.3   & 1 \\
$*M14$  & 18-03-30 & 11:02-12:18 & 23.6 & 121.9 $\pm$ 3.5   & 2 \\
\hline
$M15$   & 18-04-02 & 09:45-11:00 & 23.6 & 145.5 $\pm$ 4.8   & 2 \\
$M16$   & 18-04-02 & 11:02-12:18 & 23.6 & 171.6 $\pm$ 4.1   & 2 \\
\hline
$M17$   & 18-04-06 & 10:34-11:50 & 23.6 & 163.8 $\pm$ 3.8   & 2 \\
$M18$   & 18-04-06 & 11:52-13:07 & 23.6 & 156.3 $\pm$ 3.5   & 1 \\
\hline
$*M19$  & 18-04-07 & 09:39-11:26 & 23.6 & 121.6 $\pm$ 3.3   & 2 \\
$M20$  & 18-04-07 & 11:28-13:14 & 23.6 & 155.0 $\pm$ 4.0   & 0 \\
\hline
$M21$   & 18-04-12 & 10:20-11:22 & 18.1 & 133.6 $\pm$ 3.1   & 3 \\
$*M22$  & 18-04-12 & 11:24-12:26 & 26.1 & 228.3 $\pm$ 7.4   & 0 \\
\hline
$M23$   & 18-04-13 & 07:30-08:39 & 18.1 & 103.5 $\pm$ 2.8   & 1 \\
$*M24$  & 18-04-13 & 08:40-09:47 & 26.1 & 175.1 $\pm$ 5.2   & 1 \\
$*M25$  & 18-04-13 & 10:00-11:09 & 22.1 & 132.8 $\pm$ 2.8   & 3 \\
$*M26$  & 18-04-13 & 11:11-12:20 & 18.1 & 100.7 $\pm$ 3.1   & 3 \\
\hline
$M27$   & 18-04-17 & 10:45-11:47 & 18.1 & 219.0 $\pm$ 8.4   & 1 \\
\hline
$M28$   & 18-04-18 & 11:20-12:29 & 18.1 & 121.2 $\pm$ 2.4   & 5 \\
$M29$   & 18-04-18 & 12:31-13:40 & 26.1 & 209.6 $\pm$ 5.1   & 1 \\
\hline
$M30$   & 18-04-19 & 09:40-10:43 & 18.1 & 128.6 $\pm$ 3.8   & 2 \\
$*M31$  & 18-04-19 & 10:45-11:48 & 26.1 & 264.2 $\pm$ 7.0   & 0 \\
\hline
$M32$   & 18-04-20 & 07:40-08:58 & 18.1 & 108.8 $\pm$ 2.4   & 2 \\
$M33$   & 18-04-20 & 09:25-10:39 & 26.1 & 217.8 $\pm$ 5.6   & 1 \\
$M34$   & 18-04-20 & 10:45-12:03 & 23.6 & 153.8 $\pm$ 3.4   & 1 \\
\hline
$M35$   & 18-04-23 & 09:31-10:48 & 18.1 & 144.3 $\pm$ 2.7   & 1 \\
$M36$   & 18-04-23 & 10:51-12:08 & 18.1 & 70.3 $\pm$ 1.1    & 1 \\
\hline
$M37$   & 18-04-25 & 10:45-12:02 & 18.1 & 112.9 $\pm$ 2.1   & 1 \\
\hline
$M38$   & 18-04-26 & 09:00-10:17 & 18.1 & 122.5 $\pm$ 2.9   & 1 \\
\hline
$M39$   & 18-06-10 & 10:15-11:30 & 18.1 & 126.2 $\pm$ 3.2   & 2 \\
$M40$   & 18-06-10 & 11:32-12:47 & 26.1 & 204.1 $\pm$ 7.2   & 1 \\
\hline
$M41$   & 18-06-17 & 08:56-10:10 & 18.3 & 90.0 $\pm$ 2.6    & 1 \\
$M42$   & 18-06-17 & 10:12-11:27 & 26.1 & 258.7 $\pm$ 5.6   & 0 \\
\hline
$M43$   & 18-06-23 & 09:00-10:15 & 18.3 & 169.6 $\pm$ 3.2   & 3 \\
$*M44$  & 18-06-23 & 10:17-11:31 & 26.1 & 247.7 $\pm$ 4.0   & 4 \\
\hline
$M45$   & 18-08-06 & 10:00-11:15 & 18.3 & 131.9 $\pm$ 3.3   & 3 \\
$M46$   & 18-08-06 & 11:18-12:32 & 26.1 & 202.5 $\pm$ 4.1   & 2 \\
\hline
$M47$   & 18-08-20 & 09:50-11:05 & 18.3 & 143.5 $\pm$ 2.5   & 3 \\
$M48$   & 18-08-20 & 11:08-12:22 & 26.1 & 251.0 $\pm$ 4.7   & 1 \\
\hline
$M49$   & 18-09-21 & 10:00-11:15 & 18.3 & 110.4 $\pm$ 3.6   & 1 \\
$M50$   & 18-09-21 & 11:18-12:32 & 26.1 & 192.3 $\pm$ 6.4   & 0 \\
\hline
$M51$   & 18-09-27 & 10:20-11:35 & 18.3 & 115.8 $\pm$ 3.8   & 2 \\
$M52$   & 18-09-27 & 11:38-12:52 & 26.1 & 123.4 $\pm$ 3.8   & 0 \\
\hline
$M53$   & 18-10-03 & 09:30-10:45 & 18.3 & 120.3 $\pm$ 2.5   & 2 \\
$M54$   & 18-10-03 & 10:48-12:02 & 26.1 & 179.6 $\pm$ 4.3   & 2 \\
\hline
$M55$   & 18-10-11 & 10:40-11:55 & 18.3 & 111.1 $\pm$ 3.2   & 1 \\
$*M56$  & 18-10-11 & 11:57-13:12 & 18.3 & 195.7 $\pm$ 6.2   & 0 \\
\hline
$M57$   & 18-10-22 & 11:35-12:50 & 18.3 & 97.8 $\pm$ 2.6    & 2 \\
$*M58$  & 18-10-22 & 12:52-14:07 & 26.1 & 155.0 $\pm$ 4.7   & 0 \\
\hline
$*M59$  & 18-10-27 & 09:50-11:02 & 18.3 & 146.3 $\pm$ 4.6   & 0 \\
\hline
$M60$   & 18-10-31 & 09:45-10:59 & 18.3 & 98.9 $\pm$ 3.1    & 1 \\
$*M61$  & 18-10-31 & 11:02-12:16 & 26.1 & 190.5 $\pm$ 5.5   & 1 \\
\hline
$M62$   & 18-11-09 & 10:29-11:40 & 18.3 & 89.7 $\pm$ 2.7    & 1 \\
$*M63$  & 18-11-09 & 11:42-12:57 & 26.1 & 147.8 $\pm$ 4.2   & 1 \\
\hline
$M64$   & 18-11-12 & 10:07-11:20 & 18.3 & 109.5 $\pm$ 3.2   & 2 \\
$M65$   & 18-11-12 & 11:23-12:37 & 26.1 & 177.6 $\pm$ 5.9   & 1 \\
\hline
$M66$   & 18-11-26 & 10:05-11:20 & 18.3 & 108.3 $\pm$ 3.7   & 1 \\
$*M67$  & 18-11-26 & 11:23-12:37 & 26.1 & 162.5 $\pm$ 4.9   & 1 \\
\hline
$M68$   & 18-11-29 & 10:05-11:20 & 18.3 & 93.4 $\pm$ 2.9    & 1 \\
$M69$   & 18-11-29 & 11:23-12:37 & 26.1 & 127.2 $\pm$ 4.1   & 0 \\
\hline
$M70$   & 18-12-06 & 11:15-12:30 & 18.3 & 104.2 $\pm$ 3.2   & 1 \\
$*M71$  & 18-12-06 & 12:33-13:47 & 26.1 & 348.5 $\pm$ 13.2  & 1 \\
\hline
$S72$   & 19-05-17 & 08:40-10:25 & 25.5 & 113.0 $\pm$ 1.0   & 1 \\
$S73$   & 19-05-17 & 10:25-12:10 & 25.5 & 101.7 $\pm$ 0.9   & 1 \\
\hline
$M74$   & 19-06-12 & 08:30-09:45 & 18.3 & 113.9 $\pm$ 2.3   & 2 \\
$*M75$  & 19-06-12 & 09:47-11:02 & 25.8 & 245.3 $\pm$ 6.1   & 0 \\
\hline
$M76$   & 19-06-13 & 13:15-14:30 & 18.3 & 86.1 $\pm$ 2.9    & 2 \\
$M77$   & 19-06-13 & 14:33-15:47 & 25.8 & 158.6 $\pm$ 5.1   & 1 \\
\hline
$M78$   & 19-06-20 & 08:45-10:00 & 18.3 & 92.5 $\pm$ 3.2    & 2 \\
$M79$   & 19-06-20 & 10:03-11:17 & 25.8 & 218.8 $\pm$ 6.5   & 0 \\
\hline
$M80$   & 19-06-26 & 08:30-09:45 & 18.3 & 78.1 $\pm$ 2.9    & 1 \\
$M81$   & 19-06-26 & 09:48-11:02 & 25.8 & 196.9 $\pm$ 6.5   & 0 \\
\hline
$M82$   & 19-07-03 & 08:30-09:45 & 18.3 & 110. $\pm$ 2.7    & 3 \\
$*M83$  & 19-07-03 & 09:47-11:02 & 25.8 & 226.7 $\pm$ 5.5   & 1 \\
\hline
$M84$   & 19-07-04 & 09:05-10:20 & 18.3 & 143.1 $\pm$ 2.9   & 3 \\
$*M85$  & 19-07-04 & 10:22-11:37 & 25.8 & 207.6 $\pm$ 3.7   & 2 \\
\hline
$M86$   & 19-07-12 & 09:15-10:30 & 18.3 & 97.4 $\pm$ 2.2    & 1 \\
$*M87$  & 19-07-12 & 10:32-11:47 & 25.8 & 184.4 $\pm$ 5.4   & 0 \\
\hline
$M88$   & 19-07-19 & 08:00-09:15 & 18.3 & 92.2 $\pm$ 2.3    & 1 \\
$*M89$  & 19-07-19 & 09:17-10:32 & 25.8 & 161.4 $\pm$ 4.9   & 1 \\
\hline
$M90$   & 19-07-24 & 09:35-10:50 & 18.3 & 92.1 $\pm$ 2.4    & 2 \\
$*M91$  & 19-07-24 & 10:52-12:07 & 25.8 & 196.7 $\pm$ 5.1   & 0 \\
\hline
$M92$   & 19-07-30 & 08:10-09:25 & 18.3 & 126.7 $\pm$ 2.7   & 1 \\
$M93$   & 19-07-30 & 09:27-10:42 & 25.8 & 195.1 $\pm$ 5.5   & 1 \\
\hline
$M94$   & 19-08-07 & 10:00-11:14 & 18.3 & 85.4 $\pm$ 2.2    & 2 \\
$M95$   & 19-08-07 & 11:17-12:31 & 25.8 & 248.5 $\pm$ 5.0   & 1 \\
\hline
$M96$   & 19-08-13 & 10:00-11:15 & 18.3 & 140.1 $\pm$ 4.0   & 0 \\
$M97$   & 19-08-13 & 11:17-12:32 & 25.8 & 175.3 $\pm$ 6.6   & 0 \\
\hline
$M98$   & 19-08-27 & 09:15-10:30 & 18.3 & 110.8 $\pm$ 3.5   & 0 \\
$M99$   & 19-08-27 & 10:32-11:47 & 25.8 & 243.1 $\pm$ 8.0   & 0 \\
\hline
$M100$  & 19-09-05 & 07:30-08:45 & 18.3 & 94.1 $\pm$ 3.7    & 1 \\
$M101$  & 19-09-05 & 08:48-10:02 & 25.8 & 160.6 $\pm$ 6.0   & 0 \\
\hline
$M102$  & 19-09-10 & 09:50-11:05 & 18.3 & 88.5 $\pm$ 2.3    & 1 \\
$M103$  & 19-09-10 & 11:08-12:22 & 25.8 & 116.6 $\pm$ 3.9   & 0 \\
\hline
$M104$  & 19-09-16 & 09:10-10:25 & 18.3 & 101.8 $\pm$ 2.7   & 0 \\
$M105$  & 19-09-16 & 10:28-11:42 & 25.8 & 116.3 $\pm$ 3.7   & 0 \\
\hline
$M106$  & 19-09-24 & 10:55-12:10 & 18.3 & 81.0 $\pm$ 2.2    & 1 \\
$*M107$ & 19-09-24 & 12:13-13:27 & 25.8 & 168.1 $\pm$ 7.1   & 0 \\
\hline
$M108$  & 19-10-01 & 09:10-10:25 & 18.3 & 89.9 $\pm$ 2.7    & 1 \\
$M109$  & 19-10-01 & 10:28-11:42 & 25.8 & 120.2 $\pm$ 4.1   & 1 \\
\hline
$M110$  & 19-10-09 & 09:10-10:25 & 18.3 & 101.6 $\pm$ 3.5   & 3 \\
$M111$  & 19-10-09 & 10:28-11:42 & 25.8 & 112.2 $\pm$ 3.9   & 0 \\
$S112$  & 19-10-09 & 11:46-13:30 & 18.8 & 97.9 $\pm$ 1.0    & 3 \\
$S113$  & 19-10-09 & 09:14-10:58 & 24.7 & 130.1 $\pm$ 1.1   & 1 \\
\hline
$M114$  & 19-10-14 & 12:40-13:55 & 18.3 & 84.8 $\pm$ 2.6    & 1 \\
$M115$  & 19-10-14 & 13:58-15:12 & 25.8 & 170.1 $\pm$ 5.8   & 0 \\
\hline
$M116$  & 19-10-23 & 11:50-13:05 & 18.3 & 90.2 $\pm$ 2.5    & 1 \\
$M117$  & 19-10-23 & 13:08-14:22 & 25.8 & 123.7 $\pm$ 4.7   & 0 \\
\hline
$M118$  & 19-11-04 & 10:15-11:30 & 18.3 & 245.5 $\pm$ 9.4   & 1 \\
$*M119$ & 19-11-04 & 11:33-12:47 & 25.8 & 259.1 $\pm$ 8.7   & 0 \\
\hline
$M120$  & 19-11-14 & 10:25-11:40 & 18.3 & 82.4 $\pm$ 2.4    & 2 \\
$M121$  & 19-11-14 & 11:42-12:57 & 25.8 & 136.8 $\pm$ 4.3   & 1 \\
\hline
$M122$  & 19-11-21 & 10:10-11:25 & 18.3 & 64.9 $\pm$ 1.9    & 2 \\
$*M123$ & 19-11-21 & 11:27-12:42 & 25.8 & 129.1 $\pm$ 4.3   & 1 \\
\hline
$M124$  & 19-11-26 & 10:00-11:15 & 18.3 & 69.4 $\pm$ 2.6    & 1 \\
$*M125$ & 19-11-26 & 11:17-12:32 & 25.8 & 100.8 $\pm$ 3.9   & 0 \\
\hline
$S126$  & 20-01-28 & 09:41-11:25 & 18.8 & 162.7 $\pm$ 2.8   & 2 \\
$*S127$ & 20-01-28 & 11:26-13:18 & 18.8 & 252.7 $\pm$ 3.8   & 2 \\
$S128$  & 20-01-28 & 13:40-15:24 & 24.7 & 122.2 $\pm$ 1.0   & 3 \\
\hline
$M129$  & 20-08-10 & 09:10-10:24 & 18.3 & 83.6 $\pm$ 1.9    & 3 \\
$M130$  & 20-08-10 & 10:27-11:42 & 25.8 & 167.1 $\pm$ 5.5   & 1 \\
\hline
$M131$  & 20-08-17 & 09:15-10:29 & 18.3 & 96.8 $\pm$ 2.1    & 3 \\
$M132$  & 20-08-17 & 10:32-11:47 & 25.8 & 198.0 $\pm$ 5.5   & 2 \\
\hline
$M133$  & 20-08-26 & 09:10-10:24 & 18.3 & 97.8 $\pm$ 2.9    & 1 \\
$M134$  & 20-08-26 & 10:27-11:42 & 25.8 & 174.4 $\pm$ 5.2   & 0 \\
\hline
$M135$  & 20-09-02 & 09:15-10:29 & 18.3 & 129.2 $\pm$ 4.9   & 1 \\
$M136$  & 20-09-02 & 10:32-11:47 & 25.8 & 202.0 $\pm$ 5.4   & 0 \\
\hline
$M137$  & 20-09-06 & 08:00-09:14 & 18.3 & 93.0 $\pm$ 3.1    & 3 \\
$M138$  & 20-09-06 & 09:17-10:32 & 25.8 & 165.9 $\pm$ 5.0   & 0 \\
\hline
$M139$  & 20-09-14 & 10:10-11:24 & 18.3 & 86.7 $\pm$ 2.1    & 2 \\
$M140$  & 20-09-14 & 11:27-12:42 & 25.8 & 209.0 $\pm$ 6.0   & 0 \\
\hline
$S141$  & 20-09-17 & 09:18-11:02 & 18.8 & 141.1 $\pm$ 1.8   & 3 \\
$S142$  & 20-09-17 & 11:18-13:02 & 24.7 & 108.0 $\pm$ 1.1   & 5 \\
\hline
$M143$  & 20-09-29 & 09:00-10:14 & 18.3 & 125.5 $\pm$ 3.9   & 2 \\
$M144$  & 20-09-29 & 10:17-11:32 & 25.8 & 209.5 $\pm$ 6.0   & 2 \\
\hline
$*M145$ & 20-10-05 & 09:40-10:54 & 18.3 & 192.2 $\pm$ 5.0   & 0 \\
$*M146$ & 20-10-05 & 10:57-12:12 & 25.8 & 451.5 $\pm$ 16.5  & 0 \\
\hline
$M147$  & 20-10-19 & 08:00-09:14 & 18.3 & 78.3 $\pm$ 2.9    & 1 \\
$M148$  & 20-10-19 & 09:17-10:32 & 25.8 & 101.5 $\pm$ 3.6   & 2 \\
\hline
$S149$  & 20-10-29 & 08:32-10:05 & 24.7 & 156.3 $\pm$ 1.8   & 4 \\
$S150$  & 20-10-29 & 10:17-12:01 & 18.8 & 133.0 $\pm$ 1.6   & 2 \\
\hline
$M151$  & 20-10-30 & 08:15-09:29 & 18.3 & 67.2 $\pm$ 1.9    & 2 \\
$M152$  & 20-10-30 & 09:32-10:47 & 25.8 & 110.7 $\pm$ 3.9   & 3 \\
\hline
$M153$  & 20-11-06 & 09:52-11:07 & 25.8 & 111.1 $\pm$ 3.8   & 3 \\
$M154$  & 20-11-06 & 08:35-09:49 & 18.3 & 565.2 $\pm$ 19.4  & 1 \\
\hline
$M155$  & 20-11-08 & 09:22-10:32 & 25.8 & 123.7 $\pm$ 4.1   & 2 \\
\hline
$M156$  & 20-11-11 & 10:00-11:14 & 18.3 & 84.9 $\pm$ 2.8    & 4 \\
$M157$  & 20-11-11 & 11:17-12:32 & 25.8 & 124.0 $\pm$ 4.1   & 1 \\
\hline
$M158$  & 20-11-23 & 09:15-10:29 & 18.3 & 96.3 $\pm$ 3.7    & 4 \\
$M159$  & 20-11-23 & 12:22-13:37 & 25.8 & 184.3 $\pm$ 6.4   & 2 \\
\hline
$M160$  & 20-11-28 & 10:35-11:49 & 18.3 & 241.0 $\pm$ 10.8  & 2 \\
$M161$  & 20-11-28 & 11:52-13:04 & 25.8 & 179.4 $\pm$ 6.6   & 1 \\
\hline
$M162$  & 20-11-30 & 09:25-10:39 & 18.3 & 161.6 $\pm$ 5.3   & 3 \\
$M163$  & 20-11-30 & 10:42-11:57 & 25.8 & 120.4 $\pm$ 6.0   & 3 \\
\hline
$M164$  & 20-12-07 & 09:15-13:29 & 18.3 & 362.2 $\pm$ 15.3  & 2 \\
$*M165$ & 20-12-07 & 10:32-11:47 & 25.8 & 1029.9 $\pm$ 63.2 & 0 \\
\hline
$M166$  & 20-12-13 & 09:15-13:29 & 18.3 & 81.8 $\pm$ 2.9    & 4 \\
$M167$  & 20-12-13 & 10:32-11:47 & 25.8 & 105.0 $\pm$ 3.9   & 1 \\
\hline
$M168$  & 20-12-21 & 09:45-10:59 & 18.3 & 91.6 $\pm$ 3.9    & 4 \\
$M169$  & 20-12-21 & 11:02-12:17 & 25.8 & 120.2 $\pm$ 5.4   & 2 \\
   \hline
\end{longtable}
\begin{landscape} 
\centering
\setlength{\leftskip}{-1000pt}
\setlength{\rightskip}{\leftskip}
\scriptsize
\vspace*{\fill}
\hspace*{\fill}
\begin{longtable}{c|c|c|c|c|c|c|c|c}
\caption{Analysis results obtained through the automatic procedure of {\sc SUNDARA}.
\textit{ID} is the identification number for each single map, where the letters \textit{M} and \textit{S} specify the radiotelescope (Medicina or SRT); $ar\_id$ is the AR name (if present), \textit{Epoch} the observation date, $\nu_{obs}$ the central observing frequency, \textit{Size} the AR size, at twice the fitted semi-axes level (in units of arcmin$^2$); $T_{p,ex}$ indicates the peak of the excess brightness temperature $T_{ex}$ for each AR; $S_{sub}$ and $S_{tot}$ indicate the AR flux density of the QS-subtracted image and the original image, respectively.
\textit{Notes} indicates further AR flags: "b" indicates if the AR position is located outside of the $95\%$-level of the solar radius; "k" indicates the distance between 2 different ARs $\leq 2$~beams of the receiver; "C" indicates a AR located inside a confused region (see Footnote~\ref{footnote:conf_reg}); sequential numbers are related to multiple AR detection for the same observing session.
AR candidates without known HEK counterpart are labeled as \textit{SD\_AR\_X12345 (lon, lat)}, where \textit{SD} indicates SunDish project, \textit{AR} Active Region, \textit{X} the INAF radio telescope (\textit{M} for Medicina, \textit{S} for SRT, \textit{N} for Noto), \textit{12345} the number of AR (sorted by epoch), \textit{lon/lat} indicate the Helioprojective Longitude/Latitude (in units of arcsec).}
\label{Tab:fluxes} \\
\hline
\hline
{ID}    & $ar\_id$ & Epoch      & $\nu_{obs}$   & Size             & $T_{p,ex}$   & $S_{sub}$  & $S_{tot}$ & Notes \\
        &          & (yy-mm-dd) & (GHz) 	    & (arcmin$^2$)	   & (K)          & (sfu)      & (sfu)     &       \\
\hline
\endfirsthead
\caption{Continued.} \\
\hline
{ID}    & $ar\_id$ & Epoch      & $\nu_{obs}$   & Size             & $T_{p,ex}$   & $S_{sub}$  & $S_{tot}$ & Notes \\
        &          & (yy-mm-dd) & (GHz)         & (arcmin$^2$)     & (K)          & (sfu)      & (sfu)     &       \\
\hline
\endhead
\hline
\multicolumn{9}{l}{\footnotesize\itshape \underline{\textbf{This is a sample of the full table available on-line as Electronic Supplementary Material in Solar Physics.}}} \\
\multicolumn{9}{l}{\footnotesize\itshape $^a$ Unknown flux because the brightness temperature of the AR pixels is lower than the quiet-Sun.} \\
\endfoot
\hline
\multicolumn{9}{l}{\footnotesize\itshape \underline{\textbf{This is a sample of the full table available on-line as Electronic Supplementary Material in Solar Physics.}}} \\
\multicolumn{9}{l}{\footnotesize\itshape $^a$ Unknown flux because the brightness temperature of the AR pixels is lower than the quiet-Sun.} \\
\endlastfoot
$M2$   & NOAA$\_$12699, SPoCA$\_$21552      & 18-02-15  & 24.1 & 3.60  & $545 \pm 14$  & $0.13 \pm 0.003$ & $5.47 \pm 0.14$   & b1  \\
$M1$   & NOAA$\_$12699, SPoCA$\_$21552      & 18-02-15  & 24.1 & -     & $556 \pm 14$  & $0.17 \pm 0.004$ & $6.58 \pm 0.16$   & b1  \\
$M3$   & SD-AR$\_$M00001 (-36.2, +334.2)    & 18-02-19  & 24.1 & 49.50 & $222 \pm 6$   & $0.70 \pm 0.017$ & $74.25 \pm 1.86$  & C1  \\
$M4$   & SD-AR$\_$M00001 (-36.2, +334.2)    & 18-02-19  & 24.1 & -     & $< 284.98$    & $< 0.66$         & $< 106.31$        & 1   \\
$M5$   & NOAA$\_$12700, SPoCA$\_$21600      & 18-02-27  & 23.6 & 39.00 & $575 \pm 14$  & $0.97 \pm 0.024$ & $56.68 \pm 1.42$  & C1  \\
$M6$   & NOAA$\_$12700, SPoCA$\_$21600      & 18-02-27  & 23.6 & -     & $607 \pm 15$  & $1.75 \pm 0.044$ & $112.93 \pm 2.82$ & 1   \\
$M7$   & SPoCA$\_$21612                     & 18-03-07  & 23.6 & 51.50 & $475 \pm 12$  & $1.07 \pm 0.027$ & $74.55 \pm 1.86$  & 1   \\
$M8$   & SPoCA$\_$21612                     & 18-03-07  & 23.6 & -     & $< 216.00$    & $< 0.25$         & $< 72.70$         & 1   \\
$M10$  & SPoCA$\_$21612                     & 18-03-09  & 23.6 & 49.50 & $495 \pm 12$  & $1.14 \pm 0.028$ & $71.79 \pm 1.79$  & C1  \\
$M9$   & SPoCA$\_$21612                     & 18-03-09  & 23.6 & -     & $568 \pm 14$  & $1.22 \pm 0.030$ & $70.82 \pm 1.77$  & C1  \\
$M14$  & NOAA$\_$12703, SPoCA$\_$21684      & 18-03-30  & 23.6 & 3.75  & $360 \pm 9$   & $0.07 \pm 0.002$ & $5.41 \pm 0.14$   & b1  \\
$M13$  & NOAA$\_$12703, SPoCA$\_$21684      & 18-03-30  & 23.6 & -     & $< 314.90$    & $< 0.05$         & $< 5.03$          & 1   \\
$M15$  & SD-AR$\_$M00002 (-726.9, +16.3)    & 18-04-02  & 23.6 & 18.62 & $452 \pm 11$  & $0.44 \pm 0.011$ & $27.03 \pm 0.68$  & Ck1 \\
$M16$  & SD-AR$\_$M00002 (-726.9, +16.3)    & 18-04-02  & 23.6 & -     & $534 \pm 13$  & $0.41 \pm 0.010$ & $14.34 \pm 0.36$  & k1  \\
$M16$  & NOAA$\_$12703, SPoCA$\_$21684      & 18-04-02  & 23.6 & 49.00 & $935 \pm 23$  & $1.85 \pm 0.046$ & $71.84 \pm 1.80$  & k2  \\
$M15$  & NOAA$\_$12703, SPoCA$\_$21684      & 18-04-02  & 23.6 & -     & $763 \pm 19$  & $2.18 \pm 0.054$ & $136.50 \pm 3.41$ & k2  \\
$M17$  & NOAA$\_$12703, SPoCA$\_$21708      & 18-04-06  & 23.6 & 21.25 & $370 \pm 9$   & $0.41 \pm 0.010$ & $30.74 \pm 0.77$  & Ck1 \\
$M18$  & NOAA$\_$12703, SPoCA$\_$21708      & 18-04-06  & 23.6 & -     & $< 308.35$    & $< 0.46$         & $< 31.51$         & 1   \\
$M17$  & SD-AR$\_$M00003 (+33.7, +20.4)     & 18-04-06  & 23.6 & 47.75 & $376 \pm 9$   & $0.98 \pm 0.024$ & $69.18 \pm 1.73$  & Ck2 \\
$M18$  & SD-AR$\_$M00003 (+33.7, +20.4)     & 18-04-06  & 23.6 & -     & $383 \pm 10$  & $1.11 \pm 0.028$ & $70.38 \pm 1.76$  & Ck2 \\
$M19$  & SPoCA$\_$21710                     & 18-04-07  & 23.6 & 17.75 & $326 \pm 8$   & $0.27 \pm 0.007$ & $25.61 \pm 0.64$  & Ck1 \\
$M20$  & SPoCA$\_$21710                     & 18-04-07  & 23.6 & -     & $< 340.32$    & $< 0.34$         & $< 27.11$         & 1   \\
$M19$  & SD-AR$\_$M00004 (+243.1, -51.0)    & 18-04-07  & 23.6 & 49.25 & $377 \pm 9$   & $0.95 \pm 0.024$ & $71.28 \pm 1.78$  & Ck2 \\
$M20$  & SD-AR$\_$M00004 (+243.1, -51.0)    & 18-04-07  & 23.6 & -     & $< 338.01$    & $< 0.74$         & $< 70.37$         & 2   \\
$M21$  & SD-AR$\_$M00005 (+496.9, -24.1)    & 18-04-12  & 18.1 & 19.25 & $305 \pm 8$   & $0.18 \pm 0.005$ & $16.80 \pm 0.42$  & Ck1 \\
$M22$  & SD-AR$\_$M00005 (+496.9, -24.1)    & 18-04-12  & 26.1 & -     & -$^a$         & -$^a$            & $< 31.29$         & 1   \\
$M21$  & NOAA$\_$12704, SPoCA$\_$21724      & 18-04-12  & 18.1 & 6.00  & $797 \pm 20$  & $0.19 \pm 0.005$ & $5.37 \pm 0.13$   & bk2 \\
$M22$  & NOAA$\_$12704, SPoCA$\_$21724      & 18-04-12  & 26.1 & -     & $< 49.93$     & -$^a$            & $< 10.16$         & 2   \\
$M21$  & SD-AR$\_$M00006 (-472.4, +298.1)   & 18-04-12  & 18.1 & 42.50 & $349 \pm 9$   & $0.52 \pm 0.013$ & $37.21 \pm 0.93$  & k3  \\
$M22$  & SD-AR$\_$M00006 (-472.4, +298.1)   & 18-04-12  & 26.1 & -     & $< 138.96$    & $< 0.13$         & $< 72.92$         & 3   \\
...    & ...                                & ...       & ...          & ...           & ...              & ...               & ... \\
\end{longtable}
\hspace*{\fill}
\vfill
\end{landscape}
\begin{landscape}
\centering
\setlength{\leftskip}{-1000pt}
\setlength{\rightskip}{\leftskip}
\scriptsize
\vspace*{\fill}
\hspace*{\fill}
\begin{longtable}{c|c|c|c|c|c|c|c|c|c}
\caption{Spectral indices obtained through the automatic procedure of {\sc SUNDARA}.
$\alpha_{T_p}$, $\alpha_{T_{ex}}$, and $\alpha_{tot}$ indicate the spectral indices referred to $T_{p,tot}$, $T_{ex}$, and $S_{tot}$, respectively.
See the caption of Table~\ref{Tab:fluxes} for a full description of the other parameters.
}
\label{Tab:spec_index} \\
\hline
\hline
{ID}    & $ar\_id$ & Epoch      & T    & $\nu_{obs}$ & Size           	    & $\alpha_{T_p}$ & $\alpha_{T_{ex}}$         & $\alpha_{tot}$  & Notes \\
        &          & (yy-mm-dd) & (UT) & (GHz) 	     & (arcmin$^2$)	        &       &                                    &                 &       \\
\hline
\endfirsthead
\caption{Continued.} \\
\hline
{ID}    & $ar\_id$ & Epoch      & T    & $\nu_{obs}$ & Size           	    & $\alpha_{T_p}$ & $\alpha_{T_{ex}}$         & $\alpha_{tot}$  & Notes \\
        &          & (yy-mm-dd) & (UT) & (GHz) 	     & (arcmin$^2$)     	&       &                                    &                 &       \\
\hline
\endhead
\hline
\multicolumn{10}{l}{\footnotesize\itshape \underline{\textbf{This is a sample of the full table available on-line as Electronic Supplementary Material in Solar Physics.}}} \\
\multicolumn{10}{l}{\footnotesize\itshape $^a$ Unknown flux because the brightness temperature of the AR pixels is lower than the quiet-Sun.} \\
\endfoot
\hline
\multicolumn{10}{l}{\footnotesize\itshape \underline{\textbf{This is a sample of the full table available on-line as Electronic Supplementary Material in Solar Physics.}}} \\
\multicolumn{10}{l}{\footnotesize\itshape $^a$ Unknown flux because the brightness temperature of the AR pixels is lower than the quiet-Sun.} \\
\endlastfoot
$M21$  & SD-AR$\_$M00005 (+496.9, -24.1)    & 18-04-12 & 10:20-11:22 & 18.1 & 19.25 & $< 1.79$        & -$^a$            & $< 1.70$        & 1 \\
$M22$  & SD-AR$\_$M00005 (+496.9, -24.1)    & 18-04-12 & 11:24-12:26 & 26.1 & -     & $-$             & $-$              & $-$             & 1 \\
$M21$  & NOAA$\_$12704, SPoCA$\_$21724      & 18-04-12 & 10:20-11:22 & 18.1 & 6.00  & $< 1.70$        & $< -5.57$        & $< 1.74$        & 2 \\
$M22$  & NOAA$\_$12704, SPoCA$\_$21724      & 18-04-12 & 11:24-12:26 & 26.1 & -     & $-$             & $-$              & $-$             & 2 \\
$M21$  & SD-AR$\_$M00006 (-472.4, +298.1)   & 18-04-12 & 10:20-11:22 & 18.1 & 42.50 & $< 1.84$        & $< -0.52$        & $< 1.84$        & 3 \\
$M22$  & SD-AR$\_$M00006 (-472.4, +298.1)   & 18-04-12 & 11:24-12:26 & 26.1 & -     & $-$             & $-$              & $-$             & 3 \\
...    & ...                                & ...      & ...         & ...  & ...   & ...             & ...              & ...             & ... \\
\end{longtable}
\hspace*{\fill}
\vfill
\end{landscape}

\bibliographystyle{spr-mp-sola}
\bibliography{sola_bibliography}

\end{article} 
\end{document}